\documentclass{aastex631}

\usepackage{amsmath}
\usepackage{threeparttable}
\usepackage{subfigure}
\usepackage{graphicx}

\begin{document}

\title{A 50 s quasi-periodic oscillation in the early X-ray afterglow of GRB 220711B}

\author{H. Gao}
\affiliation{Institute for Frontier in Astronomy and Astrophysics, Beijing Normal University, Beijing 102206, China}
\affiliation{School of Physics and Astronomy, Beijing Normal University, Beijing 100875, China; gaohe@bnu.edu.cn}

\author{W.-H. Lei}
\affiliation{Department of Astronomy, School of Physics, Huazhong University of Science and Technology, Wuhan  430074, China}

\author{S. Xiao}
\affiliation{Guizhou Provincial Key Laboratory of Radio Astronomy and Data Processing, Guizhou Normal University, Guiyang 550001, China}
\affiliation{School of Physics and Electronic Science, Guizhou Normal University, Guiyang 550001, China}

\author{Z.-P. Zhu}
\affiliation{Department of Astronomy, School of Physics, Huazhong University of Science and Technology, Wuhan  430074, China}
\affiliation{Key Laboratory of Space Astronomy and Technology, National Astronomical Observatories, Chinese Academy of Sciences, Beijing 100012, China}

\author{L. Lan}
\affiliation{Institute for Frontier in Astronomy and Astrophysics, Beijing Normal University, Beijing 102206, China}
\affiliation{Department of Astronomy, Beijing Normal University, Beijing 100875, China}

\author{S.-K. Ai}
\affiliation{School of Physics and Technology, Wuhan University, Wuhan 430072, China}

\author{A. Li}
\affiliation{Institute for Frontier in Astronomy and Astrophysics, Beijing Normal University, Beijing 102206, China}
\affiliation{School of Physics and Astronomy, Beijing Normal University, Beijing 100875, China}

\author{N. Xu}
\affiliation{Institute for Frontier in Astronomy and Astrophysics, Beijing Normal University, Beijing 102206, China}
\affiliation{School of Physics and Astronomy, Beijing Normal University, Beijing 100875, China}

\author{T.-C. Wang}
\affiliation{Institute for Frontier in Astronomy and Astrophysics, Beijing Normal University, Beijing 102206, China}
\affiliation{School of Physics and Astronomy, Beijing Normal University, Beijing 100875, China}

\author{B. Zhang}
\affiliation{Nevada Center for Astrophysics, University of Nevada Las Vegas, NV 89154, USA}
\affiliation{Department of Physics and Astronomy, University of Nevada Las Vegas, NV 89154, USA}

\author{D. Xu}
\affiliation{Key Laboratory of Space Astronomy and Technology, National Astronomical Observatories, Chinese Academy of Sciences, Beijing 100012, China}

\author{J. P. U. Fynbo}
\affiliation{The Cosmic Dawn Centre (DAWN), Niels Bohr Institute, University of Copenhagen, Lyngbyvej 2, 2100, Copenhagen, Denmark}

\author{K. E. Heintz}
\affiliation{The Cosmic Dawn Centre (DAWN), Niels Bohr Institute, University of Copenhagen, Lyngbyvej 2, 2100, Copenhagen, Denmark}
\affiliation{Centre for Astrophysics and Cosmology, Science Institute, University of Iceland, Dunhagi 5, 107 Reykjavík, Iceland}

\author{P. Jakobsson}
\affiliation{Centre for Astrophysics and Cosmology, Science Institute, University of Iceland, Dunhagi 5, 107 Reykjavík, Iceland}

\author{D. A. Kann}
\affiliation{Instituto de Astrofísica de Andalucía (IAA-CSIC), Glorieta de la Astronomía s/n, 18008 Granada, Spain}

\author{S.-Y. Fu}
\affiliation{Key Laboratory of Space Astronomy and Technology, National Astronomical Observatories, Chinese Academy of Sciences, Beijing 100012, China}

\author{S.-Q. Jiang}
\affiliation{Key Laboratory of Space Astronomy and Technology, National Astronomical Observatories, Chinese Academy of Sciences, Beijing 100012, China}

\author{X. Liu}
\affiliation{Key Laboratory of Cosmic Rays, Ministry of Education, Tibet University, Lhasa, Tibet 850000, China}
\affiliation{Key Laboratory of Space Astronomy and Technology, National Astronomical Observatories, Chinese Academy of Sciences, Beijing 100012, China}

\author{S.-L. Xiong}
\affiliation{Key Laboratory of Particle Astrophysics, Institute of High Energy Physics, Chinese Academy of Sciences, 19B Yuquan Road, Beijing 100049, China}

\author{W.-X. Peng}
\affiliation{Key Laboratory of Particle Astrophysics, Institute of High Energy Physics, Chinese Academy of Sciences, 19B Yuquan Road, Beijing 100049, China}

\author{X.-B. Li}
\affiliation{Key Laboratory of Particle Astrophysics, Institute of High Energy Physics, Chinese Academy of Sciences, 19B Yuquan Road, Beijing 100049, China}

\author{W.-C. Xue}
\affiliation{Key Laboratory of Particle Astrophysics, Institute of High Energy Physics, Chinese Academy of Sciences, 19B Yuquan Road, Beijing 100049, China}

\date{\today}

\begin{abstract}
It is generally believed that long duration gamma-ray bursts (GRBs) originate from the core collapse of rapidly spinning massive stars and at least some of them are powered by hyper-accreting black holes. However, definite proofs about the progenitor and central engine of these GRBs have not been directly observed in the past. Here we report the existence of a Quasi-Periodic Oscillation (QPO) signature with periodic frequency $\sim$0.02 Hz in the early X-ray afterglow phase of GRB 220711B. Such a low-frequency QPO likely signals the precession of a relativistic jet launched from a GRB hyper-accreting black hole central engine. The energy injection signature from the \textbf{late} X-ray observations (from $5\times 10^2s\sim 1\times10^4s$) is consistent with the precession hypothesis. The prompt $\gamma$-ray light curve does not show any QPO signature, suggesting that the X-ray flaring emission in the early afterglow phase and prompt emission likely originate from different accretion processess, indicating that the progenitor stars of GRBs have a core-envelope structure with a stratified angular momentum distribution and the late-time accretion disk likely has a misalignment with respect to the rotation axis of the black hole. Such a misalignment is not expected in a canonical collapsar model. As a result, the QPO signature in GRB 220711B may reveal a new formation channel of long GRBs, possibly a stellar-merger-induced core collapse, with the orbital angular momentum of the binary misaligned with the spin axis of the collapsing star.

\end{abstract}

\keywords{QPO, gamma-ray burst}

\section{Introduction}
 
Gamma-ray bursts (GRBs), the most intense explosive phenomena in the universe, have garnered extensive attention for half a century since their discovery \citep{Zhang2018}. Based on their bimodal distribution in the duration-hardness diagram \citep{1993ApJ...413L.101K}, GRBs are categorized into two types: long-duration, soft-spectrum GRBs (LGRBs) and short-duration, hard-spectrum GRBs (SGRBs). These two classes are believed to originate from different progenitors, with LGRBs stemming from the collapse of Wolf–Rayet stars \citep{1993ApJ...405..273W, 1998ApJ...494L..45P, 1999ApJ...524..262M, 2006ARA&A..44..507W} and SGRBs from the mergers of compact stellar objects, such as NS-NS and NS-BH systems \citep{1986ApJ...308L..43P, 1989Natur.340..126E, 1991AcA....41..257P, 1992ApJ...395L..83N, 2017PhRvL.119p1101A}. Despite the distinct progenitors, the central engines of both GRB types share similarities, primarily featuring a rotating black hole surrounded by a rapidly hyper-accreting torus of debris. This system powers an ultra-relativistic outflow, generating both the prompt gamma rays and the subsequent afterglows in lower energy bands. In this case, any misalignment between the black hole's spin axis and the angular momentum axis of the BH-disk system can lead to precession due to the Lense–Thirring torque \citep{1918PhyZ...19..156L}. This precession can affect the ultrarelativistic jet launched from the central engine, driven either by neutrino annihilation or the Blandford–Znajek mechanism \citep{2007A&A...468..563L, 2010A&A...516A..16L, 2012ChPhB..21f9801L}.

The potential modulation of GRB light curves and spectral evolution by jet precession has been a topic of discussion for many years \citep{1996ApJ...473L..79B, 1999A&AS..138..507F, 1999ApJ...520..666P, 2001MNRAS.328..951P, 2006ChJAS...6a.342F, 2022ApJ...931L...2W}. However, direct observational evidence has remained elusive throughout the search. Recently, some GRBs have been discovered to exhibit repeating emission episodes in their light curves, with similar spectral evolution behavior among these episodes \citep{2022ApJ...931....4L}. \cite{2023ApJ...945...17G} suggested that the repeating light-curve properties of these GRBs could be interpreted within the framework of the jet precession model, provided that the detectable period within each precession cycle is shorter than the precession period itself, and the precession period is comparable to the duration of jet emission. Yet, since these bursts only contain two repeating emission episodes each, lacking the typical quasi-periodic oscillation (QPO) behavior in the light curve predicted by the precession model, it is difficult to conclusively determine that the repeating emission episodes definitely originate from jet precession. Another possible explanation is that these sources are candidates for millilensing events, although definitive evidence is lacking due to the absence of redshift measurements and angular offset detections for any of these candidates \citep{2021ApJ...918L..34W,2021ApJ...921L..29Y,2021NatAs...5..560P,2022ApJ...931....4L}.


In 2022, however, a unique GRB, 220711B, was observed. Its was triggered and monitored by various detectors, We processed the public data using standard analysis techniques (see Section \ref{data_reduce}). Although GRB 220711B does not exhibit repeating episodes during the prompt emission phase, it presents a series of QPO-like X-ray flares clearly in the early afterglow (See in Figure \ref{Swift}). In this study, we conduct a detailed multi-band analysis to explore its observational features and provide physical interpretations.

\section{Data reduction and analysis}
\label{data_reduce}

GRB 220711B was first detected by the Burst Alert Telescope (BAT) aboard the \textit{Neil Gehrels Swift Observatory} (Swift) at $T_0=18:16:28$ UT on 11 July 2022. The burst was also observed by the Gamma-ray Burst Monitor (GBM) on \textit{Fermi}. Swift's X-ray Telescope (XRT) began monitoring the BAT field 93 seconds after the trigger, identifying a bright, previously uncatalogued X-ray source at RA=$17^h 28^m 4.56^s$ and DEC=$+24^\circ 40' 51.5''$ (J2000) with an uncertainty of $2.1''$ \citep{2022GCN.32366....1D}. The Nordic Optical Telescope (NOT) followed up 2.95 hours later \citep{2022GCN.32377....1M}, detecting the source in the Sloan-\textit{z} band but not in the Sloan-\textit{r} band. Based on these observations and further analysis, GRB 220711B is likely at a high redshift ($z>1.5$). In this section, we present the data grouped by wavelength and provide a detailed analysis of the results. To facilitate subsequent discussions, we first introduce the observations in the optical band and the determination of redshift, the analysis to the observation in gamma-ray band and X-ray band would be introduced afterwards.

\subsection{Optical data analysis and photometric redshift estimation}

\begin{figure*}
\gridline{\fig{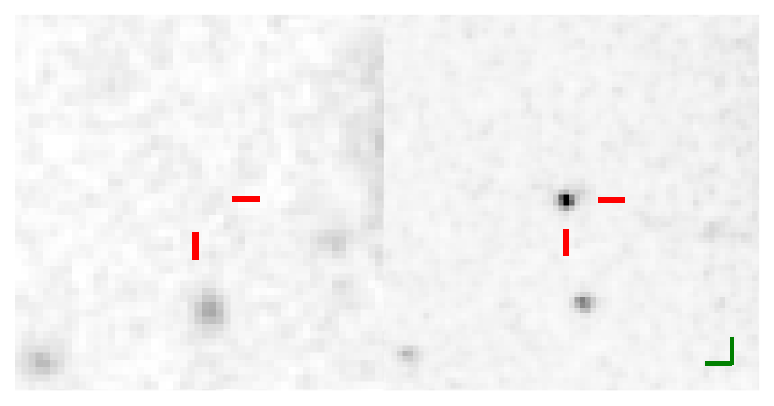}{0.76\textwidth}{(a)}}
\gridline{\fig{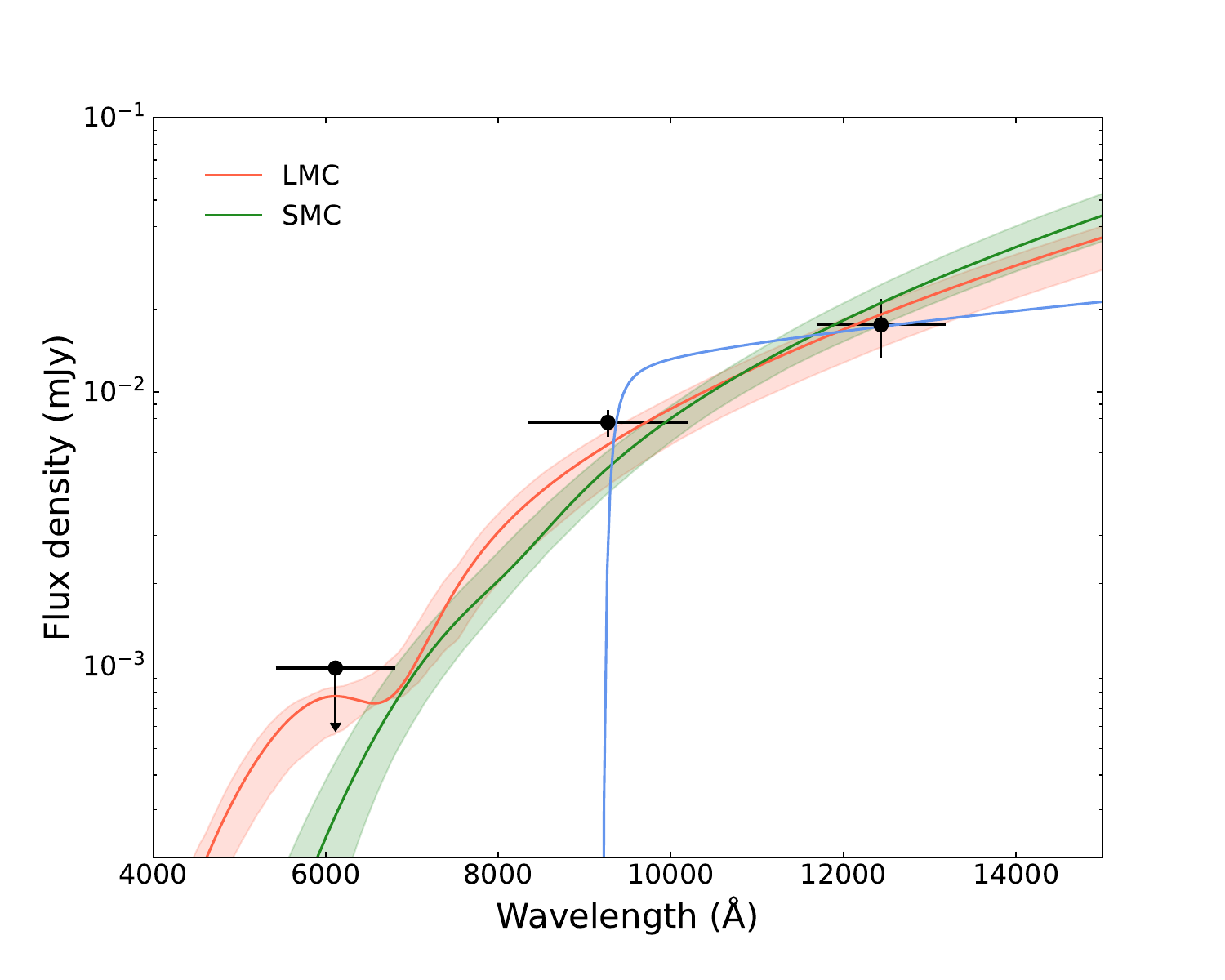}{0.48\textwidth}{(b)}
          \fig{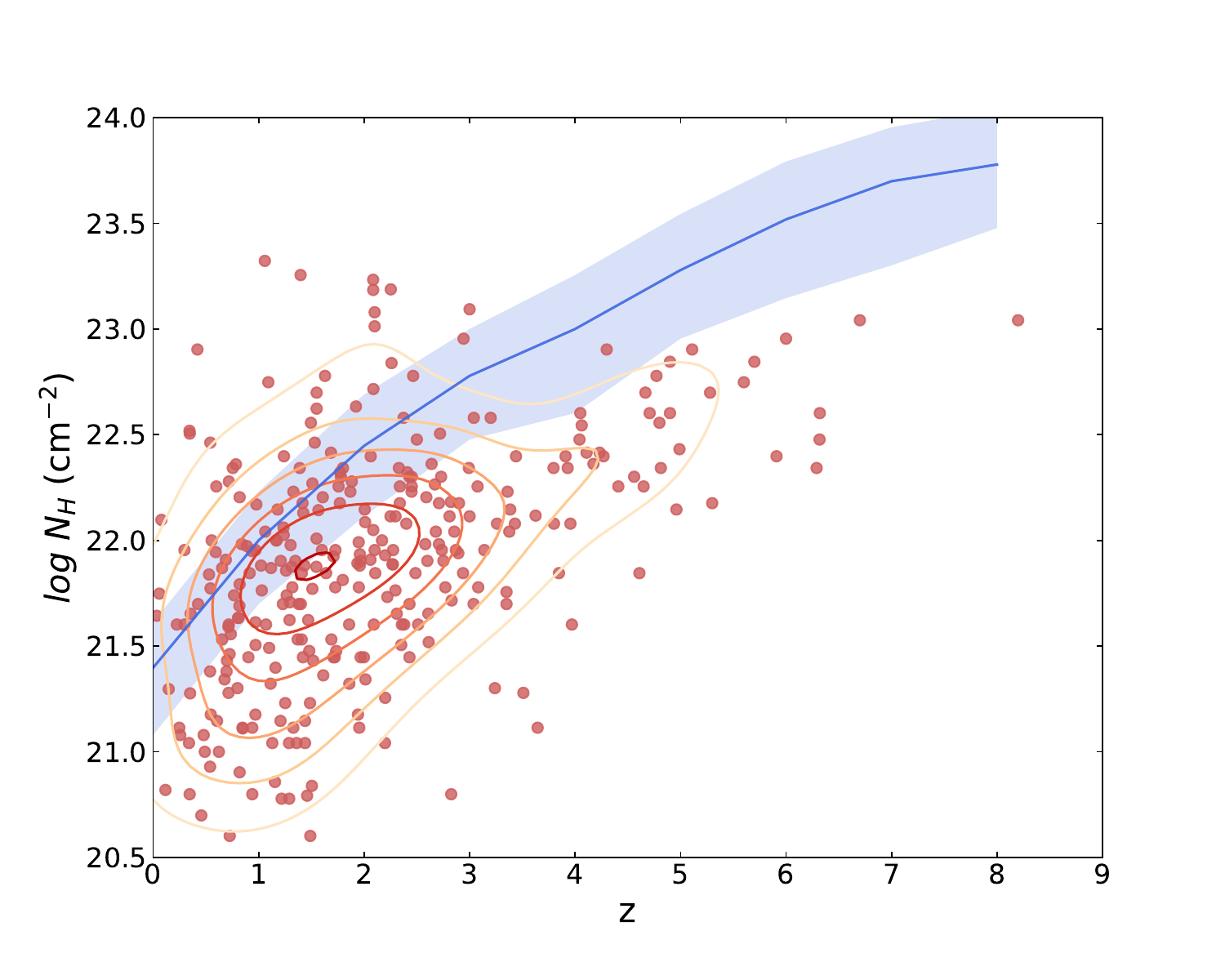}{0.48\textwidth}{(c)}}
\caption{(a) The $15^{\prime\prime}\times15^{\prime\prime}$ field view of GRB 220711B, which is at the center of the NOT Sloan-\textit{z} band image (right panel, obtained at 4.42 hr after the burst trigger), compared with the Legacy Survey archiving image in Sloan-\textit{z} band (left panel)\citep{Dey2019}. North is up and East is to the left. (b) The spectral energy distribution of GRB 220711B. Here we assume an intrinsic power-law spectrum, $F_{\nu} \propto \nu^{-\beta}$  with $\beta$ being fixed to 1.1. With no host extinction, the best fitting result is shown in blue line at a redshift of $6.52$. Additionally, assuming the host galaxy extinction follows the Large Magellanic Cloud (LMC) or Small Magellanic Cloud (SMC) dust extinction law with the extinction value $A_{V}$ being fixed to $2.45$, we get the best fit redshift are $z \simeq 2.08 \pm0.32$ and $z \simeq 2.72 \pm 0.46$, which are shown in red line and green line, respectively. The shaded region represents the $1\sigma$ confidence level. (c) The X-ray derived hydrogen column density $N_{H}$ of GRB 220711B at a redshift range from 0 to 8 (blue line) compared to GRB samples (dots in red represent individual bursts and orange lines represent their distribution contours) taken from the UK Swift Science Data Centre with known redshifts\cite{2009MNRAS.397.1177E}. The light blue shaded area represents the $1\sigma$ confidence level of $N_{H}$. It is clear that the $N_H$ value inferred from the X-ray data of GRB 220711B is less supportive to the assumption of very high redshift (e.g. $z>3$).}
\label{lo_ex_H}
\end{figure*}

\begin{figure*}
\includegraphics[width=1.0\textwidth]{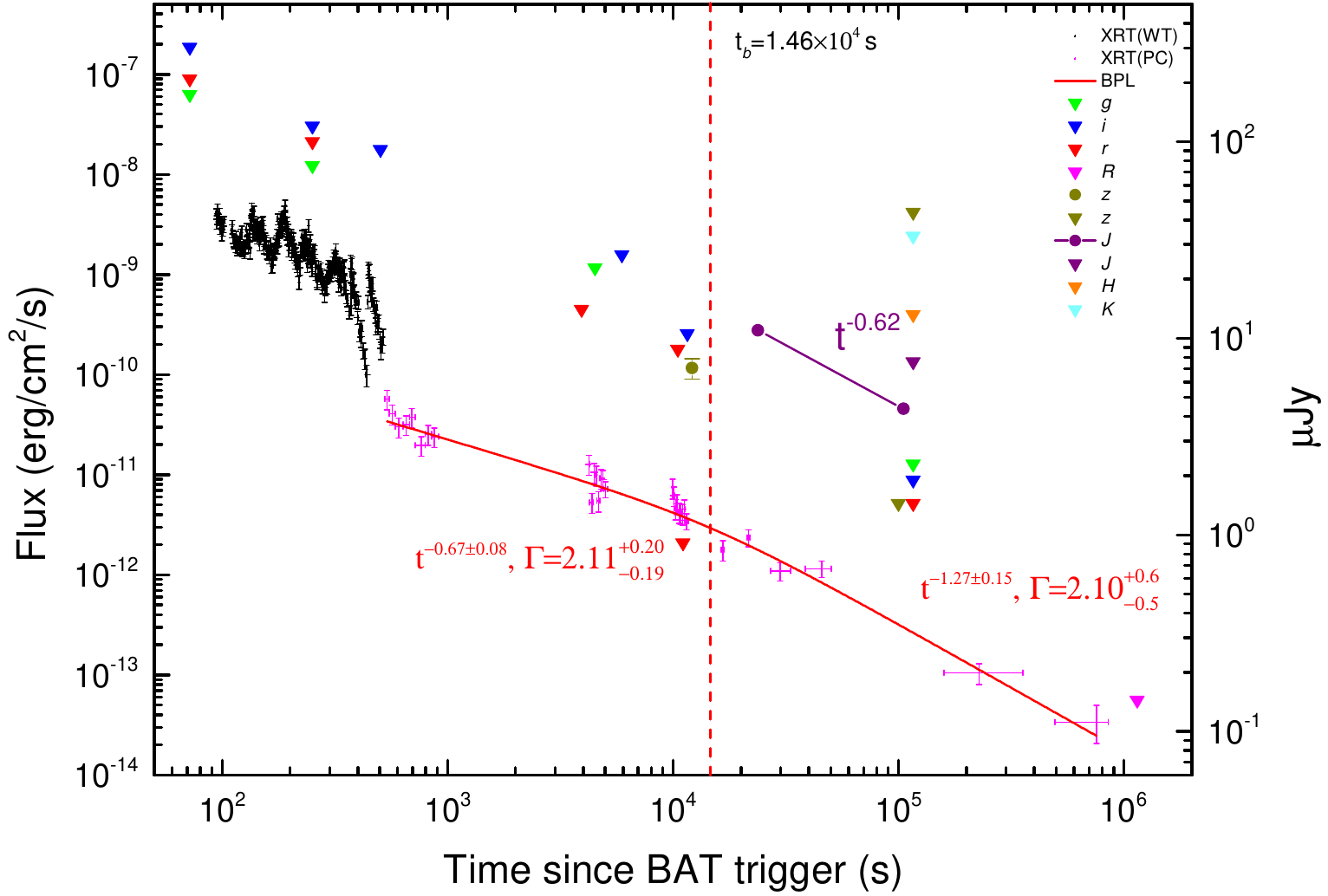}
\caption{The multi-wavelength observations and empirical fit for GRB 220711B. The observed X-ray light curves for the Windowed Timing (WT) mode and the Photon Counting (PC) mode are shown with black circles and magenta circles, respectively. The red solid line shows the broken power-law fit for the X-ray light curve in the PC mode, and the red dashed line marks the break time of the fit. The detections and upper limits of the multi-wavelength optical/NIR data are shown with circles and inverted triangles, respectively. The purple solid line shows the decay slope in the J band.}
\label{XRT}
\end{figure*}

The Nordic Optical Telescope (NOT) equipped with the Alhambra Faint Object Spectrograph and Camera (ALFOSC) started observations 2.95 hr after the BAT trigger \citep{2022GCN.32366....1D}. During the burst night, 
we obtained Sloan \emph{r} and Sloan \emph{z} band images with total exposure times of 900 s and 1000 s, respectively.
About 1.6 hours later, we triggered another 1500 s total exposure in the Sloan \emph{z} band.
Examining the stacked images with standard processing using the $\it IRAF$ package \citep{tody1986}, we found an uncatalogued source in both rounds of the \emph{z} band but undetected in the \emph{r} band, which is shown in Figure \ref{lo_ex_H} sub-figure (a) and later shown to be the optical afterglow of the burst.  We triggered VLT/FORS2 observations 13.25 days after the burst and did not detect an underlying host galaxy, deriving an upper limit of 26.0 mag in \emph{R} band. The photometric results of our observations and the public results from the Gamma-ray Coordinates Network (GCN) are listed in Table \ref{tab:phot}. The optical light curve of the burst (see Figure \ref{XRT}) can be fitted by a single power-law function with decay index $\alpha > 0.70$ for \emph{z} band and $\alpha \sim 0.62$ for \emph{J} band. It must be noted that the decay index of the J-band may be very different from the real value. The slight mismatch of decay indices may be due to the relatively large uncertainty in the \emph{J} band. 

According to the current observations, GRB host galaxies with $R_{\rm AB}>26$ mag all have redshift $z >1.5$ \citep{2009ApJ...691..182S}. Thus, 1.5 is considered to be the lower limit on the redshift for GRB 220711B.  The magnitude difference between the upper limit derived from the \emph{r} band stacked image and the detection in the \emph{z} band is larger than 2 mag during the first night of NOT observations. There are two possibilities to interpret the \emph{r} - \emph{z} drop: 1) the Lyman-break lying between the \emph{r} and \emph{z} band due to high redshift; 2) severe host galaxy reddening.

For the first scenario, the redshift needs to be in the range of $4.0\sim6.6$, in order to shift the Lyman-break to lie between the \emph{r} and \emph{z} bands. The best fit result for the observed SED (\emph{r}, \emph{z} and \emph{J} band without correction for host extinction) is obtained when $z\simeq6.52$ (see Figure \ref{lo_ex_H} sub-figure (b)). 
However, comparing with the current column density ($N_{H_{\rm X}}$) distribution from the high redshift sample (data collected from the UK Swift Science Data Centre \citep{2009MNRAS.397.1177E}), we note that the $N_{H_{\rm X}}$ value inferred from the X-ray data of GRB 220711B suggests that it is less likely to be a GRB at a very high redshift (see Figure \ref{lo_ex_H} sub-figure (c)).
 
On the other hand, we also fit the Galactic extinction corrected data of \emph{r}, \emph{z} and \emph{J} band by assuming an intrinsic power-law spectrum, $F_{\nu} \propto \nu^{-\beta}$ ($\beta$ is fixed to 1.1 according to the previous analysis) and taking $z$ as a free parameter to adjust the $\rm Ly\alpha$ absorption effect. Here we assume the host galaxy extinction follows the Large Magellanic Cloud (LMC) or Small Magellanic Cloud (SMC) dust extinction law \citep{1992ApJ...395..130P}. The extinction value $A_{V}$ is fixed to be $2.45$, which is obtained based on the relationship between $N_{HI}$ and dust extinction $A_{V}$ of GRBs \citep{2011A&A...533A..16W}, and a typical column density of $\rm{log} (\it{N}_{HI} / \rm cm^{-2}) = 21.6$ (the Gaussian peak value of 77 LGRBs \citep{2017RSOS....470304S}) is adopted for GRB 220711B. With the Monte Carlo method, we find that the best fitting result requires $z \simeq 2.08 \pm 0.32$ and $z \simeq 2.72 \pm 0.46$ for LMC extinction model and SMC extinction model, respectively (see Figure \ref{lo_ex_H} sub-figure (b)). Based on these assumptions, the \emph{r} - \emph{z} drop due to host galaxy reddening made a redshift range from $2\sim 3$. However, due to the uncertainty of the $\beta$ and $A_{V}$, we recommend a redshift of 1.5 as the safer lower limit.

\begin{center}
\begin{table*}
\centering
\caption{The photometric results of optical/NIR observations. $\Delta T$ is the median time of the exposure after the BAT trigger. Magnitudes are not corrected for Galactic extinction, which is E(B-V) = 0.07 mag \cite{2011ApJ...737..103S}.}
\label{tab:phot}
\begin{tabular}{ccccc}
\hline \hline
 $\Delta T$(hours) & Telescope/Instrument & Filter  &  Mag(AB) & Ref.   \\ 
 \hline
0.02 & BSTI & g & $>$18.3 & (2) \\
0.02 & BSTI & i & $>$17.7 & (2) \\
0.02 & BSTI & r & $>$18.1 & (2) \\
0.07 & BSTI & g & $>$19.2 & (2) \\
0.07 & BSTI & i & $>$18.7 & (2) \\
0.07 & BSTI & r & $>$18.9 & (2) \\
0.14 & GIT & i & $>$19 & (3) \\
1.09 & GIT & r & $>$21.03 & (3) \\
1.25 & GIT & g & $>$20.50 & (3) \\
1.65 & GIT & i & $>$20.34 & (3) \\
2.91 & LCO & r & $>$21.54 & (4) \\
3.08 & NOT & r & $>$24.0 & (1) \\
3.21 & LCO & i & $>$21.34 & (4) \\
3.37 & NOT & z & 21.78$\pm$0.13 & (1) \\
5.43 & NOT & z & 21.96$\pm$0.14 & (1) \\
6.61 & GTC & J & $\sim$21.3 & (5) \\
27.71 & NOT & z & $>$23.5 & (1) \\
29.25 & GTC & J & $\sim$22.3 & (5) \\
32.28 & GROND & g & $>$23.0 & (6) \\
32.28 & GROND & r & $>$23.5 & (6) \\
32.28 & GROND & i & $>$23.2 & (6) \\
32.28 & GROND & z & $>$19.8 & (6) \\
32.28 & GROND & J & $>$21.7 & (6) \\
32.28 & GROND & H & $>$21.1 & (6) \\
32.28 & GROND & K & $>$20.1 & (6) \\
318.0 & FORS2 & R & $>$26.0 & (1) \\
\hline\hline
\end{tabular}

\begin{tablenotes}
\item References: (1) this work, (2) \cite{2022GCN.32381....1K}, (3)  \cite{2022GCN.32380....1S}, (4) \cite{2022GCN.32367....1S}, (5)  \cite{2022GCN.32386....1S}, (6) \cite{2022GCN.32383....1N}
\end{tablenotes}
\end{table*}
\end{center}

\clearpage
\begin{center}
\begin{table*}
\centering
\renewcommand\tabcolsep{1.2pt}
\renewcommand\arraystretch{1.0}
\caption{Spectral fitting results in each time interval of the GRB 220711B prompt emission}
\begin{tabular}{cccc} 
\hline\hline
$t_{1}$ (s)~~~& $t_{2}$ (s) & $\alpha$ & $E_{\rm p}$ (keV)  \\
\hline
-42.4 &	-27.3  &$0.32\pm0.19$    &   $59\pm12$\\
-27.3 &	-19.7  &$-0.67\pm0.37$   &	 $73\pm11$\\
-19.7 &	-12.2  &$-0.63\pm0.27$   &	 $55\pm15$\\
-12.2 &	-2.8   &$-0.69\pm0.31$   &	 $57\pm16$\\
-2.8  &	6.5    &$-0.75\pm0.18$   &	 $140\pm18$\\
6.5  &	11.4   &$-0.61\pm0.32$   &	 $96\pm15$\\
11.4  &	20.8   &$-0.77\pm0.28$   &	 $164\pm37$\\
20.8  &	34.2   &$0.02\pm0.98$   &	 $88\pm23$\\
34.2  &	49.3   &$-0.53\pm0.21$   &	 $107\pm37$\\
49.3  &	56.9   &$-0.50\pm0.57$   &	 $85\pm19$\\
56.9  &	64.4   &$-0.62\pm0.35$   &	 $109\pm62$\\
\hline
\end{tabular}
\label{fit_data}
\end{table*}
\end{center}

\subsection{Gamma-ray data analysis}
\label{gamma_ana}

We downloaded the BAT \citep{barthelmy2005burst} data from {\it Swift} archive website
(\url{https://www.swift.ac.uk/archive/selectseq.php?source=obs&tid=1115766}). 
We downloaded the corresponding time-tagged event data of GRB 220711B from the public data site of {\it Fermi}/GBM \citep{Meegan_2009ApJ} (\url{https://heasarc.gsfc.nasa.gov/FTP/fermi/data/gbm/daily/}). 
The sodium iodide (NaI) detector, namely n8, with the smallest viewing angles in respect to the GRB source direction, was selected for our analysis. Additionally, one bismuth germanium oxide (BGO) detector, b1, closest to the GRB direction, was also selected for spectral analysis. 

\begin{figure}
\centering
\includegraphics[width=1.0\linewidth]{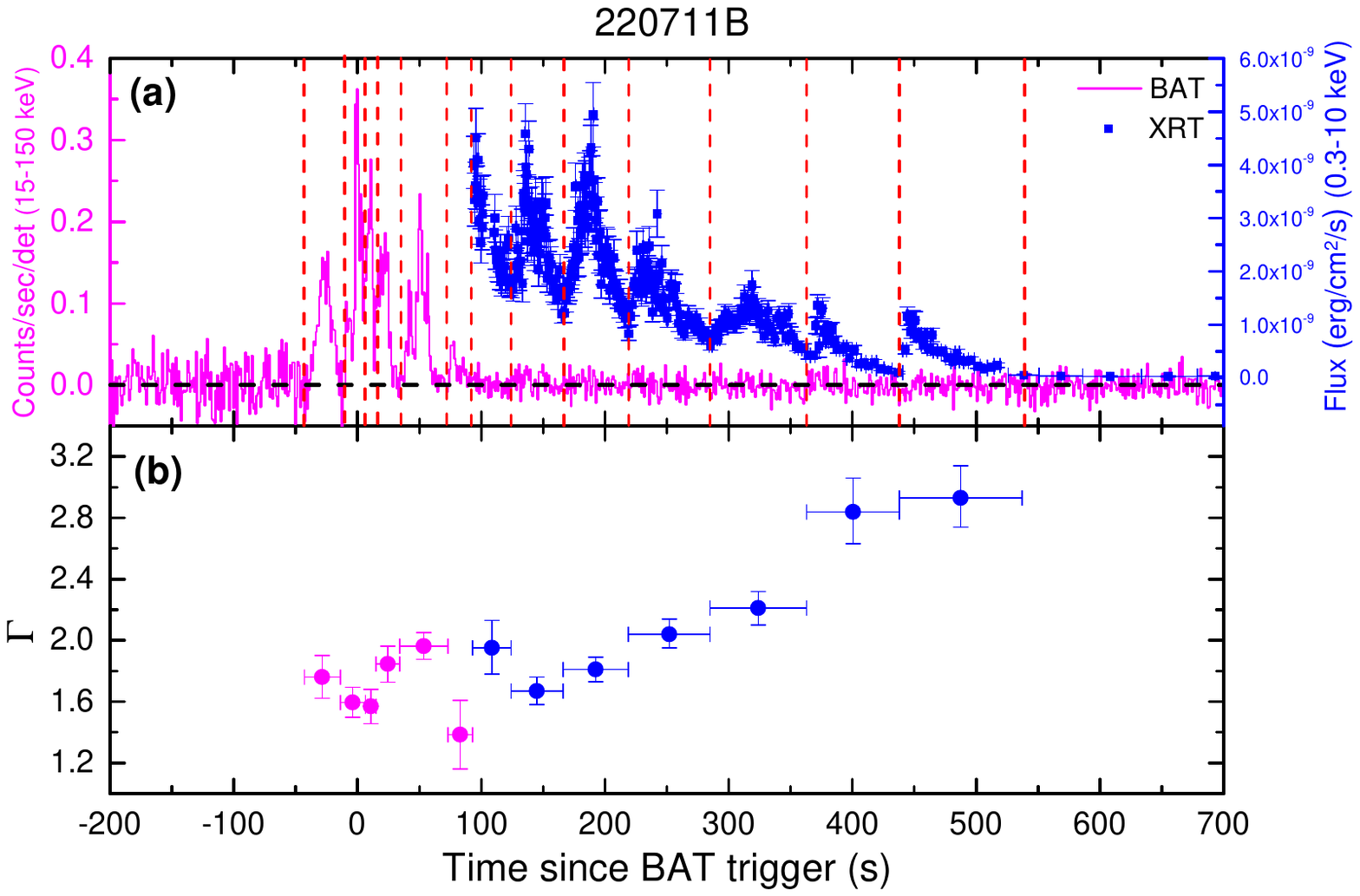}
\caption{The temporal and spectral behaviors of GRB 220711B. \textbf{a}, The light curves of BAT (magenta) and XRT (blue) obtained from {\it Swift} observations. The black horizontal dashed lines represent the BAT background flux level, and the red dashed lines show the time intervals of time-resolved spectra. \textbf{b}, The evolution of the photon spectral index ($\Gamma$) of the power-law model for the BAT data (magenta circles) and the XRT data (blue circles). All error bars represent $1-\sigma$ uncertainties.}
\label{Swift}
\end{figure}

The BAT light curve is obtained following the standard analysis threads (\url{https://www.swift.ac.uk/analysis/bat/}). We first downloaded the BAT data and used the \emph{HEASoft} tools (version 6.28) to process it, and ran the late ``convert" command from the \emph{HEASoft} software release to obtain the energy scale for the BAT events. Then the BAT light curve (Figure \ref{Swift}) is extracted with a 
bin size of 0.1 s and with the energy range of 15-350 keV using \emph{batbinevt}. The $T_{\rm 90, BAT}$ of GRB 220711B is determined by \emph{battblocks} to be $88.64\pm19.96$ s.

The GBM light curve is derived by binning the photons with a bin size of 0.064s (default bin size for GBM ctime data) in the energy range 8-40000 keV collected based on the standard \emph{HEASoft} tools (version 6.28) and the Fermi ScienceTools (v10r0p5). We choose the brightest detector among NaI and BGO to do the analysis, respectively, because the brightest detector has a minimum angle between the incident photon and the normal direction of the detector, and finally selected NaI and BGO detectors (namely, n8,b1). The background level is modeled by the baseline algorithm in the \emph{pybaseline} (\url{https://github.com/derb12/pybaselines}) package. The burst has a duration $T_{\rm 90, GBM}\sim 82$ s and is in the 10-1000 keV range.

We perform detailed time-resolved and time-integrated spectral fitting using the GBM and BAT data (see Figure \ref{Swift} and Table \ref{fit_data}). The XSPEC tool is used to perform spectral fits for each time slice, and the $\chi^2$ statistic is adopted to judge the goodness of the spectral fits. The power-law (PL), cutoff power-law (CPL) and \textbf{Band} \citep{Band_1993ApJ} functions are adopted to fit the time-resolved spectra and the time-integrated spectra, respectively. The PL, CPL and Band models can be expressed as

\begin{equation}
 N(E)=AE^{-\alpha},
\end{equation}

\begin{equation}
 N(E)=AE^{\alpha}{\rm exp}(-E/E_{\rm c}),
\end{equation}
and

\begin{equation}
 N(E)=\left\{
 \begin{array}{l}
 A(\frac{E}{100\,{\rm keV}})^{\alpha}{\rm exp}(-\frac{E}{E_{\rm c}}),\,E<(\alpha-\beta)E_{\rm c} \\
 A\big[\frac{(\alpha-\beta)E_{\rm c}}{100\,{\rm keV}}\big]^{\alpha-\beta}{\rm exp}(\beta-\alpha)(\frac{E}{100\,{\rm keV}})^{\beta}, E\geq(\alpha-\beta)E_{\rm c} \\
 \end{array}\right.
\end{equation}
respectively, where $\alpha$ and $\beta$ are low-energy and high-energy photon spectral indices. The peak energy $E_{\rm p}$ is related to the cut-off energy, $E_{\rm c}$, through $E_{\rm p}=(2+\alpha)E_{\rm c}$. Table \ref{fit_data} lists the CPL spectral fitting results for the prompt emission.
Notably, The time-integrated spectrum of the GBM data can be best fitted by a cutoff power-law model with spectral index $\alpha=-1.39^{+0.07}_{-0.07}$ and peak energy $E_{\rm p}=81^{+6}_{-6}~{\rm keV}$, the fluence within GBM energy band is $1.02^{+0.04}{-0.04}\times10^{-5}~{\rm erg/cm^2}$ calculated by the model parameters. The peak energy value is well within the reach of Swift/BAT, but that of the BAT data can be best fitted by a power-law model with photon index $\Gamma=1.75^{+0.08}_{-0.08}$. We also tried to use a cutoff power-law model to fit the BAT data, with spectral index $\alpha=-1.52^{+0.31}_{-0.31}$ and peak energy $E_{\rm p}=102^{+34}_{-34}~{\rm keV}$, the fluence within BAT band is $6.0^{+0.3}{-0.3}\times10^{-6}~{\rm erg/cm^2}$. The two $E_{\rm p}$ values are consistent with each other within the error range. 

\begin{figure*}
\gridline{\fig{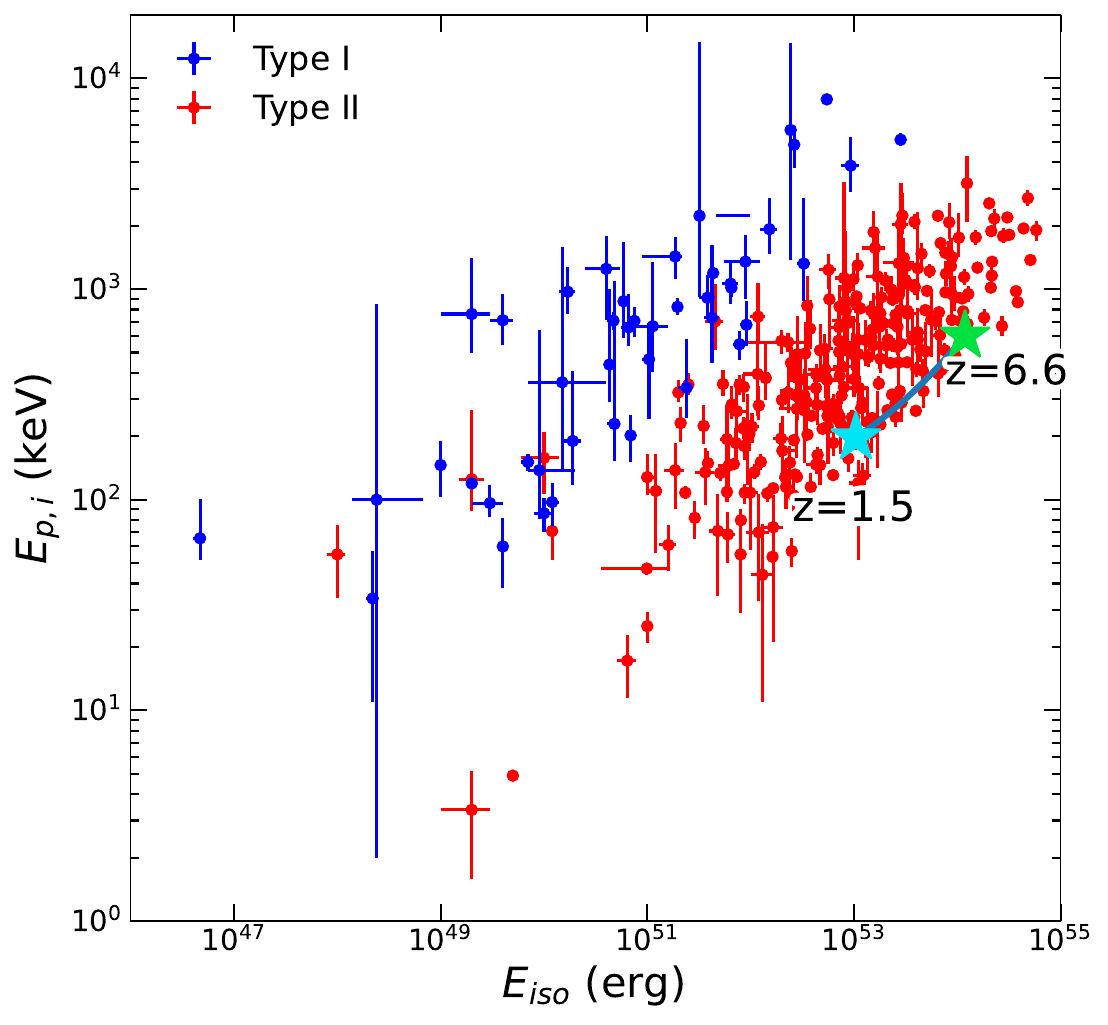}{0.47\textwidth}{(a)}
          \fig{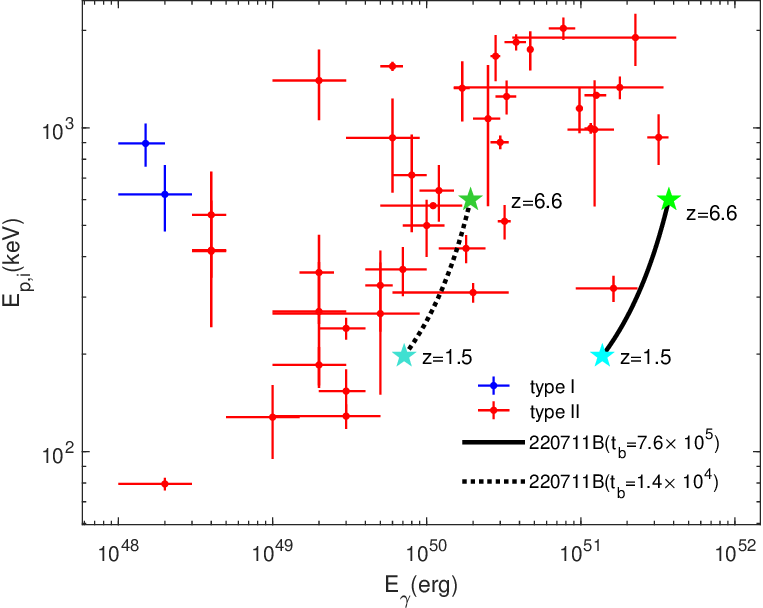}{0.53\textwidth}{(b)}}
\caption{(a) The Amati diagram with GRB 220711B (marked with cyan and green stars representing redshifts of 1.5 and 6.6, respectively). $E_{p,i}$ is the rest-frame peak energy calculated by $E_{p,i}=(1+z)E_{p}$ ($E_{p}$ is the peak energy in the $E^2N(E)$ spectrum). The blue solid line shows the position of GRB 220711B in different redshift range from 1.5 to 6.6, the $E_{p}$ of GRB 220711B is $79$ keV and its $E_{p,i}$ ranges from $197.5$ keV to $600.4$ keV. The data points for type I and type II GRBs are taken from \cite{Minaev2020}. (b) The Ghirlanda diagram with GRB 220711B. The black line shows the position of GRB 220711B in different redshift range from 1.5 to 6.6. The dotted line is for the case when $t_{\rm b}=1.4\times 10^4$ s and the solid line is for the case when $t_{\rm b}=7.6\times 10^5$ s. The data points for type I and type II GRBs are taken from \cite{wang2018}.}
\label{fig:relation}
\end{figure*}

With the estimation of redshift, we plot GRB 220711B in the Amati relation \citep{Amati_2002A&A}, which is a correlation between the GRB isotropic energy $E_{\rm\gamma,iso}$ and the rest-frame peak energy $E_{\rm p,z}=(1+z)E_{\rm p}$. Here we plot Amati diagrams (see Figure \ref{fig:relation} sub-figure (a)) using GRB samples from \cite{Minaev2020} and \cite{wang2018}. Due to the uncertainty of redshift, we show the position curve of GRB 220711B with redshift from 1.5 to 6.6. We find that in the Amati relation diagram, GRB 220711B follows the long GRB track but in the boundary region of the high energy band.

\subsection{X-ray data analysis}
\label{X_ana}

The XRT \citep{Burrows_2005SSRv} light curve (see Figure \ref{XRT}) is obtained by using of the public data from the {\it Swift} archive (\url{https://www.swift.ac.uk/xrt_curves/}). The X-ray light curve at early times (WT mode) shows a fading behavior with several superimposed regularly-spaced flares. For each flare in the WT mode data, we perform detailed time-resolved spectral fitting with similar methods to the prompt emission. The fitting results are shown in Figure \ref{Swift}. For PC mode data, we perform time-resolved spectral fitting for each segment separated by the break time $t_{\rm b}$. The best fitting result is 
$\Gamma_{\rm X,1}=2.11^{+0.20}_{-0.19}$ for $t<t_{\rm b}$, and $\Gamma_{\rm X,2}=2.10^{+0.60}_{-0.50}$ for $t>t_{\rm b}$. 

For each flare in the WT XRT light curves, the isotropic X-ray energy is calculated by $E_{\rm X,iso}=4k\pi D_{L}^2S_{\rm X}/(1+z)$, where $S_{\rm X}$ is X-ray flare fluences, which are estimated to be $2.32\times10^{-8}-1.51\times10^{-7}~\rm erg~cm^{-2}$, those fluences are 2 to 3 orders of magnitude lower than the prompt emission ($1.02\times10^{-5}~\rm erg~cm^{-2}$), $D_L$ is the GRB luminosity distance, and the k-correction factor corrects 
the XRT-band (0.3-10 keV) flux to a wide band in the burst rest frame (0.1-1000 keV in this analysis), i.e.

\begin{equation}
k=\frac{\int^{10^3/{1+z}}_{0.1/{1+z}}EN(E)dE}{\int^{10}_{0.3}EN(E)dE}.
\end{equation}
Here $N(E)$ is the time-dependent PL photon spectrum. To calculate $D_{L}(z)$, the concordance cosmology parameters
$H_0 = 67.8$ km s$^{-1}$ Mpc $^{-1}$, $\Omega_M=0.308$,and $\Omega_{\Lambda}=0.692$ have been adopted according to the {\it Planck} results \citep{Planck2018}. 

At late times (PC mode), the X-ray lightcurve seems to be a power-law $F=F_{0}t^{-\alpha}$ or broken power-law decay $F=F_{0}[(t/t_{\rm b})^{\omega \alpha_1}+(t/t_{\rm b})^{\omega \alpha_2}]^{-1/\omega}$. In order to test which model is best fitted to the late-time XRT light curve, we compared the goodness of the fits by invoking the Bayesian information criteria (BIC)\footnote{BIC is a criterion to evaluate the best-fitted model among a finite set of models, and the model with the lowest BIC is preferred. The definition of BIC can be written as: $BIC=\rm -2ln L+k\cdot ln(n)$, where $k$ is the number of model parameters, $n$ is the number of data points, and $L$ is the maximum value of the likelihood function of the estimated model. (1) if $0<\Delta BIC<2$, the evidence against the model with higher BIC is not worth more than a bare mention; (2) if $2<\Delta BIC<6$, the evidence against the model with higher BIC is positive; (3) if $6<\Delta BIC<10$, the evidence against the model with higher BIC is strong; (4) if $10<\Delta BIC$, the evidence against the model with higher BIC is very strong.}. We found that the $\Delta \rm BIC=30$ ($\rm BIC=83$ and $\rm BIC=53$ for power-law model and broken power-law model). This result suggest that the broken power-law model can be fitted to the late-time light curve better,
with the indices $\alpha_1=0.67\pm0.08$, $\alpha_2=1.27\pm0.15$, and the break time $t_{\rm b} = (1.4\pm1.2)\times10^{4}$ s. The photon spectrum indices are $\Gamma_{\rm X,1}=2.11^{+0.20}_{-0.19}$ and $\Gamma_{\rm X,2}=2.10^{+0.60}_{-0.50}$ ($N(E) \propto E^{-\Gamma}$). The spectrum with specific flux density ($F_{\nu} \propto \nu^{-\beta}$) can thus be derived as
\begin{eqnarray}
F_{\nu} \propto E N(E) \propto \nu^{-\Gamma+1} \propto \nu^{-\beta},
\end{eqnarray}
with $\beta_1 \sim 1.11^{+0.20}_{-0.19}$ and $\beta_2 \sim 1.10^{+0.60}_{-0.50}$.

Generally, the X-ray afterglow of GRB 220711B can be explained within the framework of the external shock model, while the existence of the shallow decay phase indicates that additional energy injection from the central engine should be included. There are mainly two types of energy injection mechanisms being discussed in the literature: (1) continuous injection with an evolving luminosity \citep{dai&lu1998,zhang&meszaros2001} $L_{\rm inj} = L_{\rm inj,0}(t/t_0)^{-q}$ (e.g. afterglows powered by a spinning-down NS); (2) pulsed energy injection with stratified ejecta \citep{rees&meszaros1998,sari&meszaros2000}. 
The two scenarios can be treated as equivalent, with $q$-$s$ relations introduced. 




In theory, the spectrum of the synchrotron radiation can be approximately divided into several segments with some characteristic frequencies considered. Usually, the frequency of X-ray photons is much greater than the frequency that directly related to the electrons with minimum Lorentz factor ($\nu_m$), as well as the characteristic cooling frequency ($\nu_c$). Therefore, in the shallow decay phase ($\alpha_1$), the relation between $\alpha$ and $\beta$ (closure relation) reads as \citep{gao2013}
\begin{eqnarray}
\alpha_1 = \frac{q-2}{2} + \frac{(2+q)\beta_1}{2},
\end{eqnarray}
from which $q = 0.53^{+0.33}_{-0.29}$ can be calculated. Assuming the external shock is propagating in an ISM-like ambient medium, one has \citep{zhang2006}
\begin{eqnarray}
s = \frac{10-7q}{2+q} = 2.49^{+1.22}_{-1.10}.
\end{eqnarray}

After the break, the photon density spectrum index does not change, so that the break could be due to the termination of energy injection. In this case, the post break phase is in the self-similar deceleration regime, where the closure relation reads as \citep{gao2013}
\begin{eqnarray}
\alpha = \frac{3\beta-1}{2}.
\end{eqnarray}
Obviously, considering the uncertainties, $\alpha_2$ and $\beta_2$ could satisfy the closure relation.

Alternatively, the achromatic steepening temporal break shown in the afterglow lightcurve could correspond to the jet-break effect. In this case, the relation between $\alpha$ and $\beta$ can be expressed as \citep{gao2013}
\begin{eqnarray}
\alpha = \frac{3q-2+2\beta (q+2)}{4},
\end{eqnarray}
for the ISM-like ambient medium. With $\alpha_{2}=1.27^{+0.15}_{-0.15}$ and $\beta_2 \sim 1.10^{+0.60}_{-0.50}$, we have $q = 0.52^{+0.74}_{-1.70}$ and $s = 2.52^{+2.79}_{-2.16}$, which are consistent with the values obtained from the data in the shallow decay phase. This seems to verify the hypothesis that the change of temporal index of the X-ray afterglow at $1.4 \times 10^4$ s is due to the jet break. If this is the case, the energy injection should last longer than $7.6\times10^{5}$ s, which is not easy to achieve unless the central engine keeps active for this long time or the lower end of the Lorentz factor distribution of the stratified ejecta extends to $\Gamma\sim$ a few. 

\begin{figure*}
\centering
\includegraphics[width=1.0\textwidth]{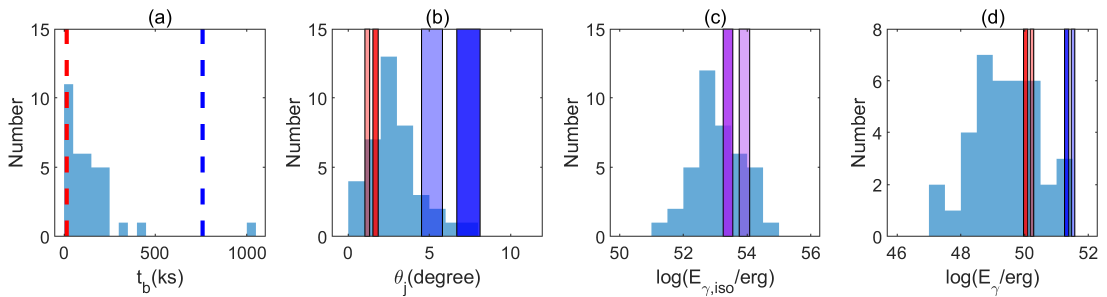}
\caption{(a) The distribution of the observed jet break time ($t_{\rm b}$). The red dashed line represents $t_{\rm b}=1.4\times 10^4$ s and blue dashed line represents $t_{\rm b}=7.6\times 10^5$ s. (b) The distribution of jet opening angle ($\theta_j$). The dark red (blue) shadowed strip marks the inferred $\theta_j$ for GRB 220711B when $t_{\rm b}=1.4\times 10^4$ s ($t_{\rm b}=7.6\times 10^5$ s) in $z=2-3$. The light red (blue) shadowed strip mark the inferred $\theta_j$ for GRB 220711B when $t_{\rm b}=1.4\times 10^4$ s ($t_{\rm b}=7.6\times 10^5$ s) in $z=4-6.6$. (c) The distribution of the isotropic $\gamma$-ray energy ($E_{\gamma,\rm iso}$). The dark purple shadowed strip represents the inferred $E_{\gamma,\rm iso}$ for  GRB 220711B when $z=2-3$. The light purple shadowed strip represents the inferred $E_{\gamma,\rm iso}$ for GRB 220711B when $z=4-6.6$. (d) The distribution of the beaming corrected $\gamma$-ray energy ($E_{\gamma}$). The dark red (blue) dashed line marks the inferred $E_{\gamma}$ for GRB 220711B when $t_{\rm b}=1.4\times 10^4$ s ($t_{\rm b}=7.6\times 10^5$ s) in $z=2-3$. The light red (blue) shadowed strip mark the inferred $E_{\gamma,iso}$ for GRB 220711B when $t_{\rm b}=1.4\times 10^4$ s ($t_{\rm b}=7.6\times 10^5$ s) in $z=4-6.6$.}
\label{fig:par_distri}
\end{figure*}

Compared with other observed sources, the possible jet break time of GRB220711B ($t_{\rm b}=1.4 \times 10^4$ s) is relatively small (see Figure \ref{fig:par_distri}). For a constant density ISM medium, the jet opening angle could be estimated as
\begin{eqnarray}
\theta_{\rm j}=0.07 {\rm rad}(\frac{t_{\rm b} }{1 {\rm day}})^{3/8}(\frac{1+z}{2})^{-3/8}(\frac{E_{\rm K,iso}}{10^{53}{\rm ergs}})^{-1/8}(\frac{n}{0.1\rm cm^{-3}})^{1/8},
\end{eqnarray}
where $n$ (here is adopted as $1~\rm cm^{-3}$) is the ISM number density and $E_{\rm K,iso}$ is the isotropic kinetic energy, which could be estimated with  
\begin{eqnarray}
\eta=\frac{E_{\gamma,\rm iso}}{E_{\gamma,\rm iso}+E_{\rm K,iso}},
\end{eqnarray}
where $\eta$ (here adopted as 0.2 \citep{zhang2007,wang2015}) is the $\gamma$-ray radiation efficiency and $E_{\gamma,\rm iso}$ is the isotropic $\gamma$-ray energy. The result of $\theta_{\rm j}$ thus depends on the redshift value. When $z=2-3$, we have $\theta_{\rm j}=1.8^{\circ}-1.5^{\circ}$, and if $z=4-6.6$, we have $\theta_{\rm j}=1.3^{\circ}-1.0^{\circ}$ (see Figure \ref{fig:par_distri}). 

If the achromatic break is due to the termination of energy injection, the jet-break time would be larger than the last X-ray detection time ($7.6\times10^{5}$ s). In this case, the result of $\theta_{\rm j}$ would be $\theta_{\rm j}=8.1^{\circ}-6.7^{\circ}$ at $z=2-3$ and $\theta_{\rm j}=5.8^{\circ}-4.5^{\circ}$ at $z=4-6.6$, which is comparable or even much larger than the maximum value for previous observations (see Figure \ref{fig:par_distri}). 

With the estimation of the jet opening angle, we plot GRB 220711B in the Ghirlanda relation \citep{ghirlanda2004}, which is a correlation between $E_{\rm p,z}$ and the beaming-corrected bolometric emission energy $E_{\gamma}$. $E_{\gamma}$ can be calculated as 
\begin{eqnarray}
E_{\gamma}=(1-\cos\theta_{\rm j})E_{\gamma,\rm iso}.
\end{eqnarray}
Here we plot both Ghirlanda diagrams (see Figure \ref{fig:relation} sub-figure (b)) using GRB samples from \cite{Minaev2020} and \cite{wang2018}. Due to the uncertainty of redshift, we show the position curve of GRB 220711B with redshift from 1.5 to 6.6. In the Ghirlanda diagram, when the achromatic break is taken as the jet-break, GRB 220711B lies in the central region of the long GRB track. Otherwise if the achromatic break is due to the termination of energy injection, GRB 220711B would deviate from other bursts due to its relatively high beaming-corrected energy.

\subsection{QPO searching}
\label{QPO}

\begin{figure}[hp!]
\centering
\includegraphics[width=0.48\textwidth]{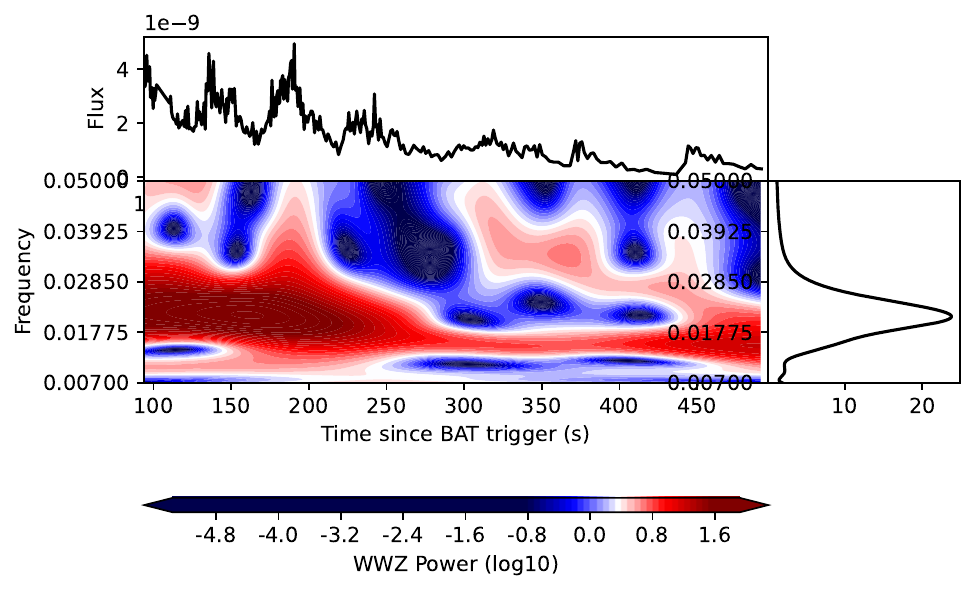}
\includegraphics[width=0.48\textwidth]{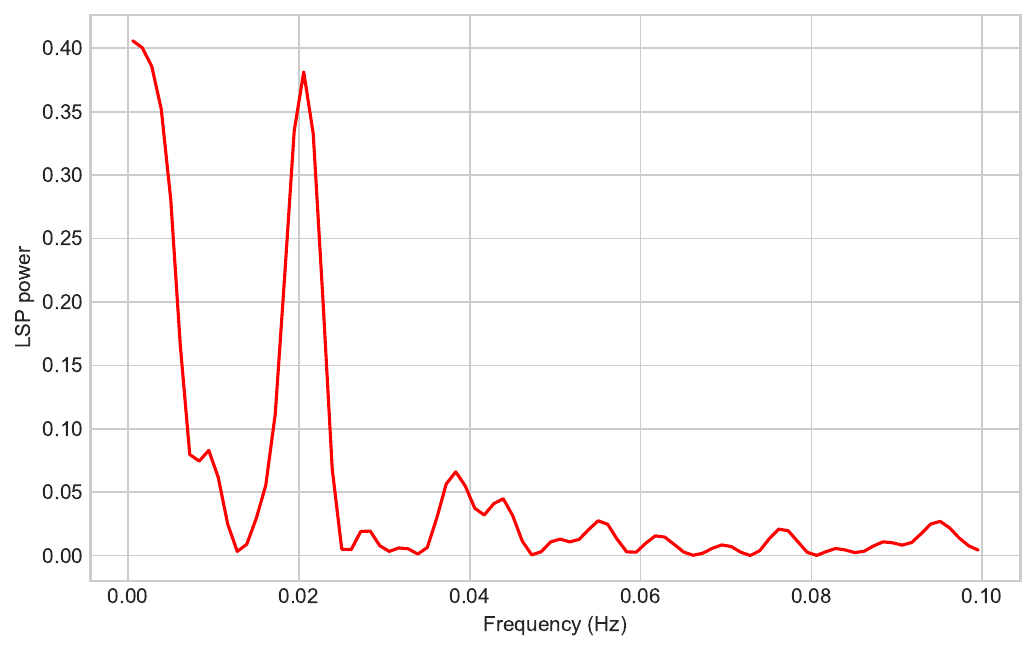}

\includegraphics[width=0.48\textwidth]{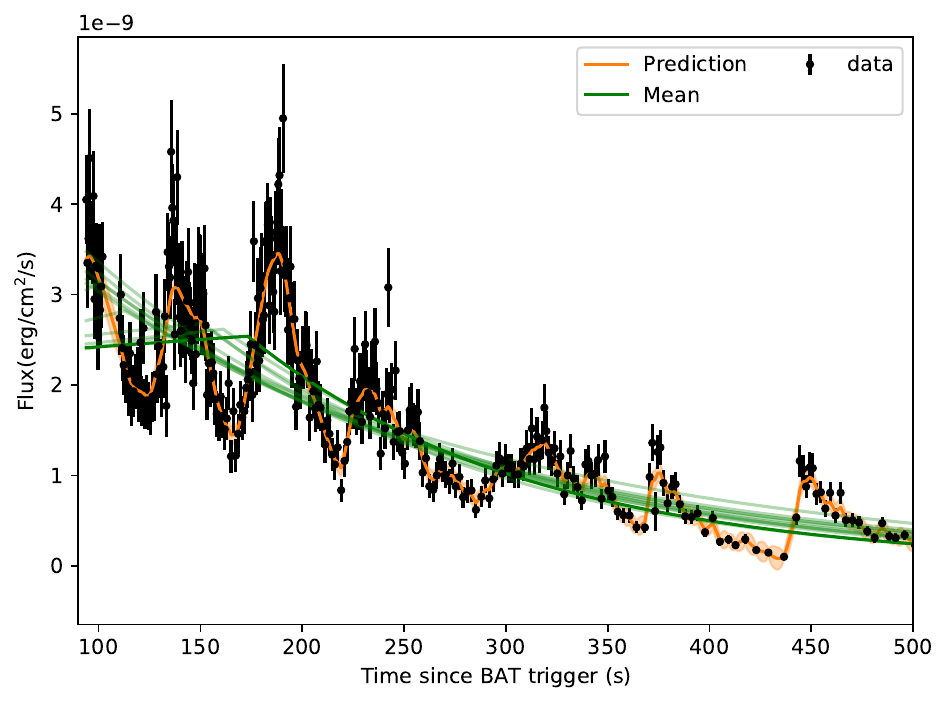}
\includegraphics[width=0.48\textwidth]{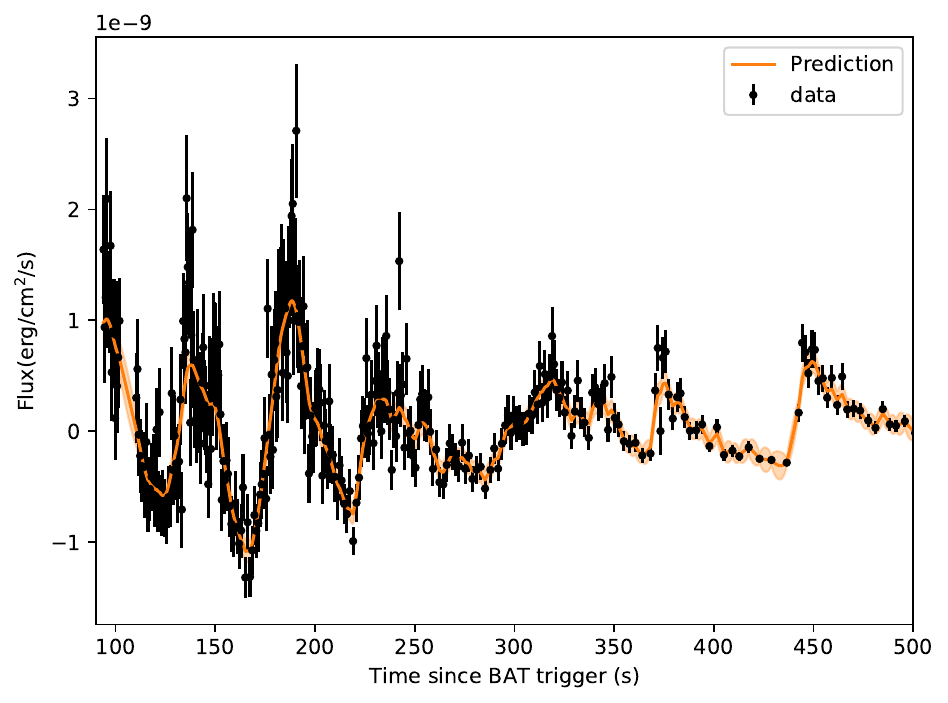}
\includegraphics[width=0.48\textwidth]{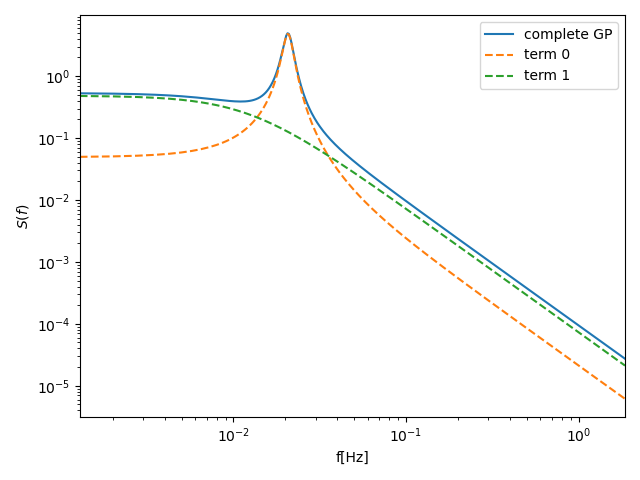}
\includegraphics[width=0.48\textwidth]{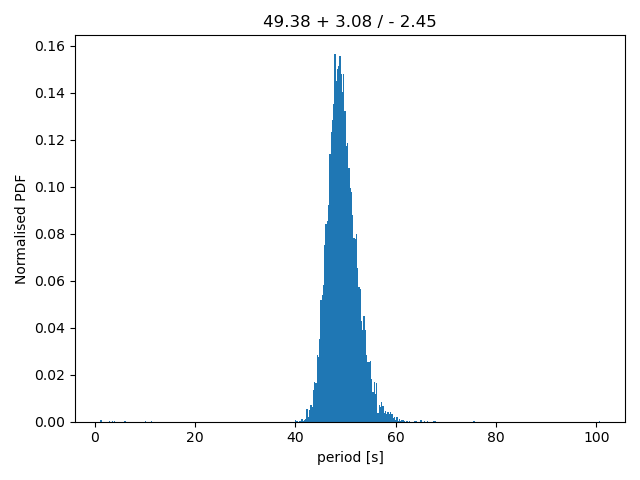}        
\caption {The Top panel are the QPO search results with the Weighted Wavelet Z-transform (WWZ) and Lomb-Scargle Periodogram (LSP) methods. Top left panel: the lightcurve of XRT and the 2D-contour plots of WWZ power in the time-frequency domain, which suggest that a significant QPO with frequency 0.02 Hz ($\sim 50 s$) exists from T0+93 s to T0+270 s, but quickly disappears after T0+270 s. Top right panel: LSP power in the frequency domain of the XRT lightcurve from T0+93 s to T0+270 s. The middle and bottom panel are the QPO search results of the XRT light curve in the time domain with the Gaussian process method. Middle pannels: the XRT lightcurve and the maximum likelihood fit (left), and the residual lightcurve and the maximum likelihood prediction (right). We adopt the procedure in \cite{hubner2022searching}, it models QPOs as a stochastic process on top of a deterministic shape (i.e. mean function or green lines). For the deterministic shape the early X-ray afterglow of GRB 220711B, we find that a exponential function (i.e. green lines) can be well fitted to it. Bottom panels: the maximum likelihood power spectra density (left), and the posterior distribution of the period (right). The $\ln BF_{\rm qpo}$ for XRT data is 8, which significantly supports the existence of a QPO \citep{hubner2022searching}.
} 
\label{GP}
\end{figure}

The Weighted Wavelet Z-transform (WWZ) \citep{foster1996wavelets} is one of the most widely used methods in QPO search in the time-frequency domain. We use this method in the XRT light curve, and the 2D-contour of WWZ power spectra suggests that the QPO with 0.02 Hz is strong from T0+93 s to T0+270 s, but quickly weakens after T0+270 s. It's worth noting here that there is a high-to-low frequency drift in the WWZ spectrum (see the Top left panel of Figure \ref{GP}).

Due to the non-uniform time sampling of the XRT data used, we adopt the Lomb-Scargle Periodogram (LSP) 
\citep{lomb1976least, scargle1982studies} method, which is another widely used method to search for QPOs in the frequency-domain. We first apply this method to the XRT light curve from T0+93 s to T0+270 s and find a strong peak at $0.02$ Hz (see the Top panel of Figure \ref{GP}) in the power spectra. 

The significance of the QPO is estimated to be 9.1 $\sigma$ by Baluev method \citep{vanderplas2018understanding}. Moreover, in order to minimize the effect of the overall afterglow profile, we also generate a Lomb-Scargle Periodogram using the exponential background subtracted light curve from T0+93 s to T0+270 s, which also shows a strong peak at $0.02$ Hz with a significance level of 13.3 $\sigma$. Here the subtracted background could be described by the mean function as $Ae^{-(t-t_s)/\tau}$, where the best fitting result is $
A=3.3_{-0.7}^{+0.9}\times10^{-9}{\rm erg~s^{-1}~cm^{-2}}$, $t_s=77_{-38}^{+78}$ s and $\tau=180_{-31}^{+31}$ s (see green lines in Figure \ref{GP}). 

\begin{figure}
\centering
\includegraphics[width=0.48\textwidth]{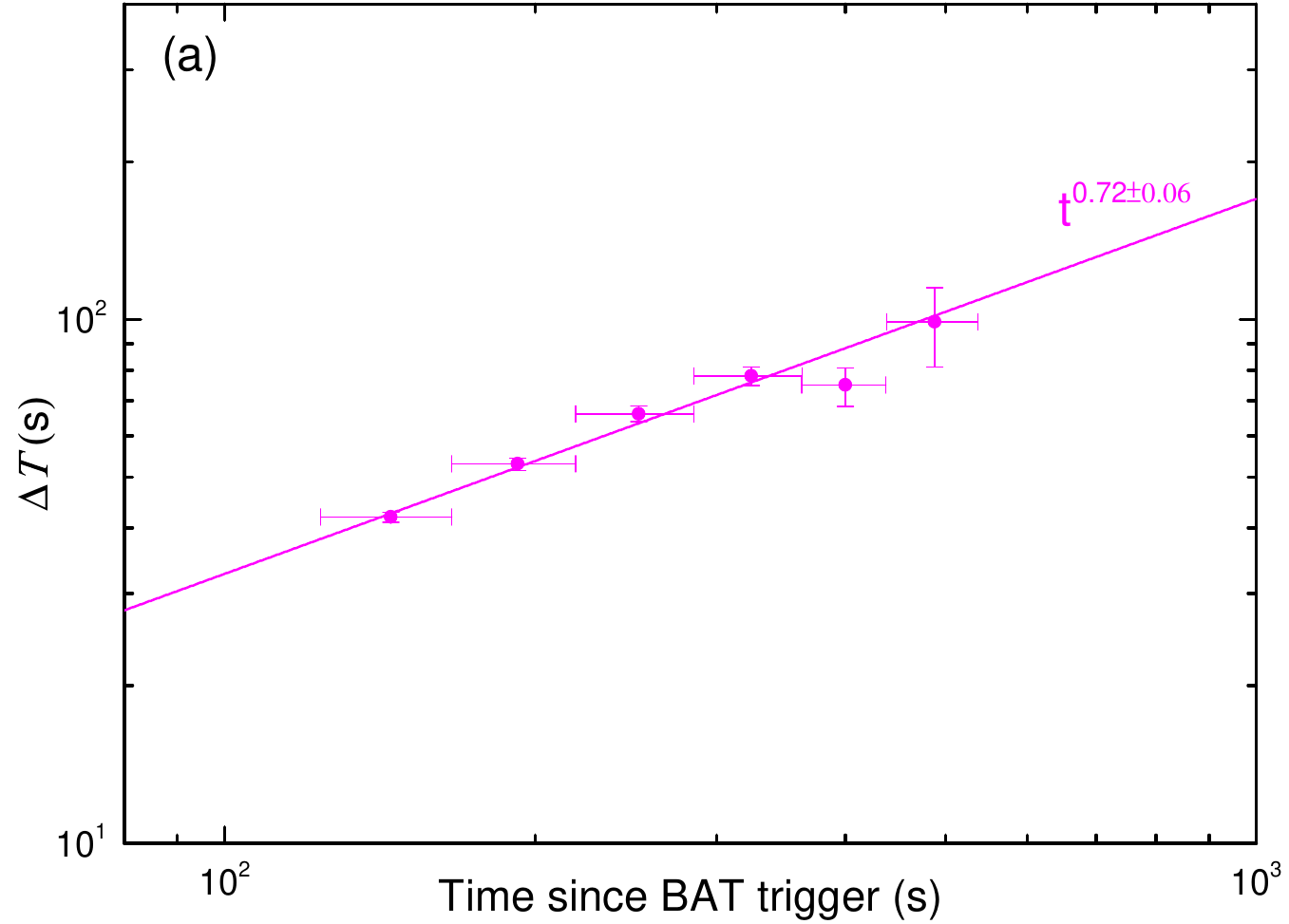}
\includegraphics[width=0.48\textwidth]{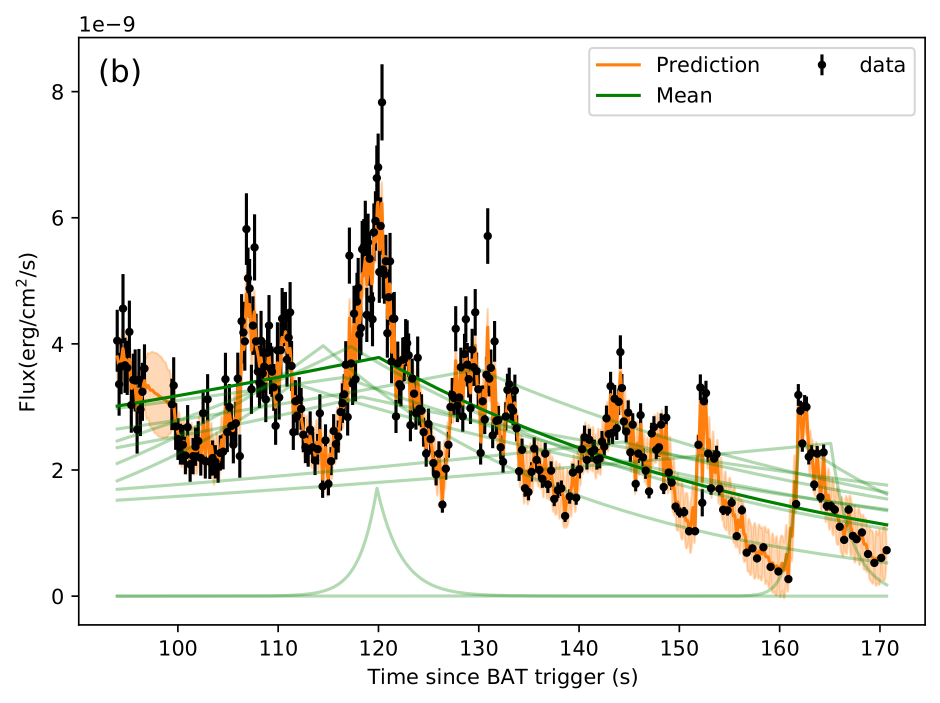}   
\includegraphics[width=0.48\textwidth]{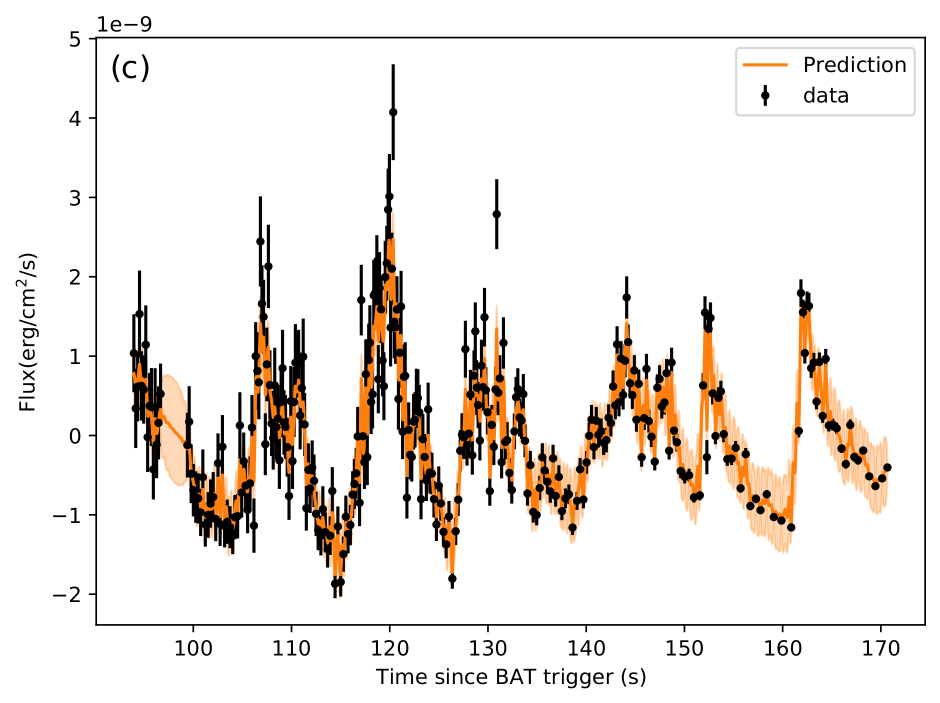}
\includegraphics[width=0.48\textwidth]{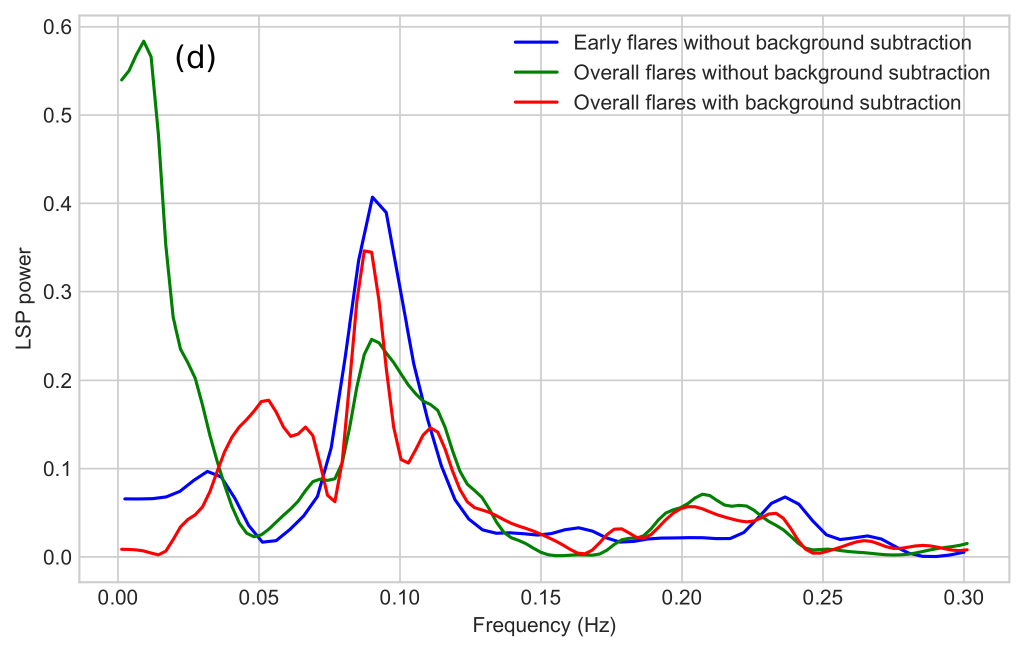}  
\caption { Top pannels: (a) the evolution of X-ray flare duration in the Windowed Timing mode of XRT light curves of GRB 220711B, where the magenta solid line shows the power-law function fit for the evolution (left). It is worth noting that when fitting the evolution pattern, we did not use the information from the first flare because the data for this flare was incomplete, making it impossible to accurately determine its width and peak time; (b) the light curve by manually deducting the overall evolution trend from the data by shrinking the time sequence with $(t/t_0)^{-0.72}$ (right). Bottom panels: the exponential background subtracted light curve (left), and the Lomb-Scargle Periodogram power of the XRT light curve of a) early flares without background subtraction (blue line); b) overall flares without background subtraction (green line); c) overall flares with background subtraction (red line) (right).} 
\label{delT}
\end{figure}

We find that the duration of each X-ray flare ($\Delta T$) gradually increases with time as $\Delta T\propto t^{0.72}$ (see Figure \ref{delT}). After manually adjusting for the overall evolution trend from the data by shrinking the time sequence with $(t/t_0)^{-0.72}$ ($t_0$ is the starting time of the XRT light curve), we conduct another QPO search with the LSP method. We find a strong QPO signal with a confidence level of $9.3~\sigma$ among the early flares that corresponds to the 93-270 s time span of the original data. For all the X-ray flares (corresponding to 93-520 s of original data), a QPO signal with a confidence level of 2.6 $\sigma$ or $\sim10$ $\sigma$ could be detected in the light curve before or after exponential background subtracting (see Figure \ref{delT}). Here the subtracted background could also be described by the mean function as $Ae^{-(t-t_s)/\tau}$, where the best fitting result is $
A=3.6_{-1.5}^{+0.7}\times10^{-9}{\rm erg~s^{-1}~cm^{-2}}$, $t_s=118_{-20}^{+10}$ s and $\tau=42_{-37}^{+32}$ s (see green lines in Figure \ref{delT}). It is worth noting that the Baluev method for estimating the significance in the LSP assumes a white noise background and may not adequately account for red noise (e.g. \cite{2005A&A...431..391V,2025A&A...693A.319K}), which may overestimate significance.

To further validate the reliability of the detected QPO signal, we employed an additional time-domain analysis method for QPO detection, namely the Gaussian Process (GP) method \citep{hubner2022searching}. In this work, the XRT light curve is modeled with QPOs as a stochastic process on top of a deterministic shape (i.e. exponential function), and  we perform model selection between QPOs and red noise (for more detailed information refer to the article \citep{hubner2022searching}). The QPO Bayes factor $BF_{\rm qpo}$ is defined as
\begin{equation}
\begin{split}\label{equ:gs}
BF_{\rm qpo}=\frac{Z(d|k_{\rm qpo+rn},\mu)}{Z(d|k_{\rm rn},\mu)},
\end{split}
\end{equation}
where the numerator (i.e. QPO and red noise) and denominator (i.e. red noise) are the respective evidences in the different models, and the $k$, $d$ and $\mu$ are kernels, data and the parameter of the mean function (i.e. exponential function), respectively. For GRB 220711B, a QPO with period $\sim$50 s is found with a GP method (see Figure \ref{GP}), with a QPO Bayes factor $\ln BF_{\rm qpo}\sim$ 8. When ${\rm In}BF_{\rm qpo}>3$, it is very unlikely that we have seen a false positive, which corresponds to a p-value of approximately 0.001 \citep{hubner2022searching}. We refer to the publicly available code of GP \footnote{\href{https://github.com/MoritzThomasHuebner/QPOEstimation}{https://github.com/MoritzThomasHuebner/QPOEstimation}} designed by \cite{hubner2022searching}.

We used Monte Carlo simulations to test the impact of the observation errors of the lightcurves on the QPO signal search results. For each time bin with a particular observed count rate, a mock count rate is generated based on the observed count rate $C$ by randomly generating the data from a normal distribution with parameters ($C$, $\sqrt{C}$). Subsequently, over 1000 mock light curves are generated by collecting these randomly generated count rates for each time bin. QPO searching methods described above are applied to each mock lightcurve to identify the QPO frequency in each realization. Our findings indicate that the lightcurve during the 93-290 s period is significantly bright, so that the observation error of the lightcurve barely affects the frequency and significance of the QPO signal.

In summary, we have identified a significant QPO signal (0.02Hz) in the early X-ray emission of GRB 220711B using various methods. We applied all three methods to the gamma-ray data of GRB 220711B, but no QPO signal was 
found. 

\section{Physical interpretation}
\label{summary}

In general, many properties of GRB 220711B indicate that it is a normal long GRB, which likely originated from the core collapse of a massive star. Such a seemingly ordinary source surprisingly contains a QPO signal that does not exist in any other GRB. Within the framework of the collapsar model, the central engine of the GRB could be either a strongly magnetized neutron star (magnetar) or a hyper-accreting black hole (BH) \citep{zhang2018book}. If the central engine is a magnetar, the QPO frequency may be related to its spin frequency or the frequency of torsional/crust oscillations. However, in order to power a GRB, the spin period of a magnetar should be of the order of milliseconds, much smaller than the QPO period of GRB 220711B. Also, according to the observations of soft Gamma-ray repeaters (SGRs), the QPOs induced by torsional/crust oscillations would be in the range 10's Hz to 1000's Hz (e.g. Giant Flares of SGR 1900+14 \citep{strohmayer2005discovery}, and the SGR J1935+2154 burst associated with a Fast Radio Burst (FRB 200428) \citep{li2022quasi}). These frequencies are  much larger than the QPO frequency of GRB 220711B. Therefore, the central engine of GRB 220711B is unlikely to be a magnetar. 

\begin{figure*}[htp]
\centering
\includegraphics[width=0.6\textwidth]{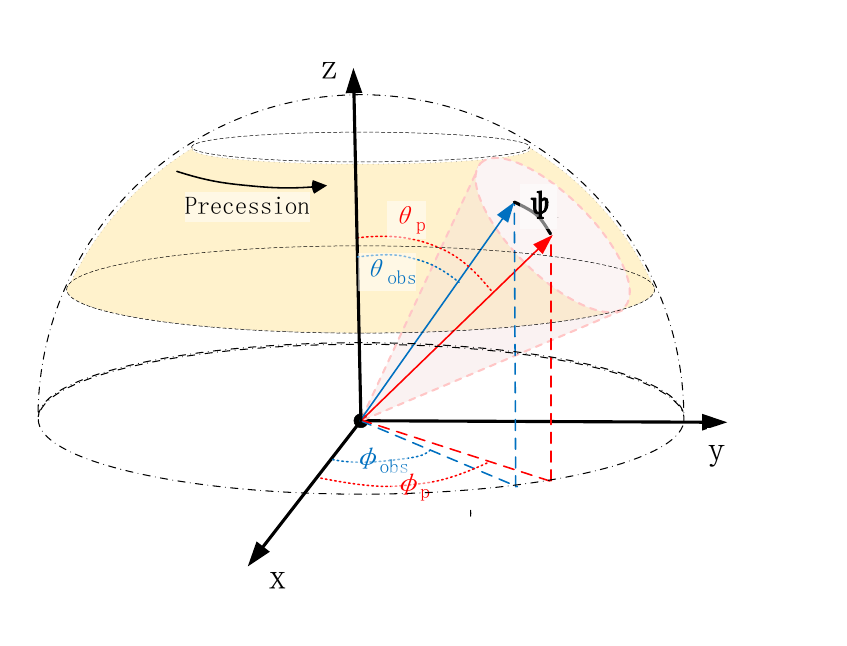}
\caption {Schematic picture of a precessing jet. $\Psi(t)$ is the angle between the observer and the central locus of the jet, $\theta_{\rm p}$ is the precession angle around the zenith, $\phi_{\rm p}$ is the azimuth angle of the jet, and the line of sight between the observer and the central object is fixed as ($\theta_{\rm obs}$,$\phi_{\rm obs}$). } \label{cartoon}
\end{figure*}

If the central engine is a hyper-accreting BH, a QPO signal could be related to the jet precession effect or the viscous instability induced by inner boundary torque (or magnetic coupling torque). The oscillating timescale for viscous instability should be in order of 100 ms \citep{Xie2016,Lei2009}, which is too short to interpret the QPO found in GRB 220711B. A precessing jet becomes the most natural interpretation. The misalignment in the spin axis of the BH and the angular momentum axis of the BH-disk system could drive the accretion disk to precess, which is known as Lense-Thirring precession \citep{1918PhyZ...19..156L}. In this case, the ultra-relativistic jet launched from the central engine would also be precessing. Due to the beaming effect, high energy photons are detectable only when $\Psi(t)\lesssim\theta_{\rm j}+1/\Gamma$, where $\Gamma$ is the Lorenz factor of the jet and $\Psi(t)$ is the angle between the observer and the central locus of the jet, which can be calculated as 

\begin{equation}
\Psi(t) = \cos ^{ - 1}\left[{\rm cos} \theta_{\rm p}
{\rm cos} \theta_{\rm obs} + {\rm sin} \theta_{\rm p} {\rm sin}
\theta_{\rm obs}{\rm cos} (\phi_{\rm p}(t) - \phi_{\rm obs})\right],
\end{equation}
where $\theta_{\rm p}$ is the precession angle around the zenith, $\phi_{\rm p}(t)=2\pi t/P$ is the azimuth angle of the jet, where $P$ is the precession period, and the line of sight between the observer and the central object is fixed as ($\theta_{\rm obs}$,$\phi_{\rm obs}$). Here we describe the schematic picture of a precessing jet in spherical coordinates (see Figure \ref{cartoon}). In every precession circle, the detectable period can be estimated as 

\begin{equation}
\Delta T=\frac{P}{\pi} \cos ^{ - 1}\left[\frac{\cos \theta_{\rm j}-\cos \theta_{\rm obs} \cos\theta_{\rm p} }{\sin \theta_{\rm obs} \sin\theta_{\rm p} }\right].
\end{equation}
The observed lightcurve profile essentially depends on the relations between $P$, $\Delta T$ and the total emission duration $T_{\rm dur}$. When $P\ll T_{\rm dur}$, multiple pulses with QPO modulation would show up. There might be intervals between pulses if $\Delta T\ll P$ or no clear intervals if $\Delta T\sim P$. GRB 220711B should belong to the situation of $\Delta T\sim P\ll T_{\rm dur}$. In this scenario, the period of the QPO is roughly the precession period $P$, which is related to the mass ($M_{\rm BH}$) and spin of the BH ($a_{*}$), the accretion rate ($\dot{M}$), and the viscosity parameter of the disc \citep{lei2013}. It is worth noting that the $\gamma$-ray signals of GRB 220711B does not contain any QPO signature (\ref{discussion}), suggesting that the $\gamma$-ray emission and the early X-ray emission should be from different accretion processes.

The QPO signature of GRB 220711B warrants a distinct interpretation from other long GRBs. We envisage the following scenario: A standard GRB progenitor (e.g. a Wolf-Rayet star) is in a close binary system with a compact companion with a smaller mass \citep{Fryer1998,Janiuk2017}. The companion merges with the inner-most core part of the progenitor star to produce a BH, and thus induces the collapse of the progenitor star. In this case, a long-lived accretion disc can be created, which could power the prompt emission of a GRB as well as its long-lived X-ray activities \citep{Barkov2010,Janiuk2013}. The duration of the prompt emission and the X-ray activities are essentially determined by the progenitor properties, such as magnetization, density and angular velocity profile of the progenitor star \citep{Gottlieb2022,Kumar2008,Kumar2008science}.

Here we assume the accretion of the outer core material of the progenitor star powers the prompt emission. Later on, a portion of the envelope materials fall back onto the BH, powering a new relativistic jet to produce flaring X-ray emission. We use the model proposed in \cite{Kumar2008,Kumar2008science} to calculate the rate of infall of stellar matter on an accretion disc during the collapse of a rapidly rotating massive star. In this model, a particle that is initially located at $(r,\theta,\phi)$ with angular velocity $\Omega(r,\theta)$ is considered to go into free-fall, colliding with its counterpart from the opposite side of the equatorial plane and intersecting the equatorial plane at

\begin{equation}
    t_{\rm{eq}} = \Omega^{-1}_{\rm{k}}[\cos^{-1}{(-e)}+e(1-e^2)^{1/2}](1+e)^{-3/2}+t_s(r),
\end{equation}
where $e=1-(\Omega/\Omega_{\rm k})^2\sin^2 \theta$ is the eccentricity, $\Omega_{\rm{k}}$ is the Keplerian angular velocity, and $t_s$ is the sound travel time from the centre to radius $r$. The initial intersect distance is $r_{\rm{eq}}(r,\theta) = r(1-e)$. The velocity field immediately following the in-fall of stellar matter on the equatorial plane is convergent toward the centre of the star, so that the radius at which the material goes into quasi-circular orbit quickly shrinks to 

\begin{equation}
    r_{\rm{fb}} = \frac{\langle r(1-e)\rangle}{1.4}.
\end{equation}
Treating the equal-collapse-time surface as spherical, the mass fall-back rate on to the equatorial plane would be 

\begin{equation}
    \dot{m}_{\rm{fb}}(t) \sim \frac{dM_r}{dr}\frac{dr}{dt_{\rm eq}}.
\end{equation}
Particles that circularise inside the innermost stable circular orbit (ISCO) of the black hole will fall into the hole on a dynamical time scale. The mass and dimensionless spin $a_*$ of the initial black hole satisfy 

\begin{equation}
R_{\rm isco}(M_{\rm{r}},a_*) = r\frac{\Omega^2(r)}{\Omega_{\rm{k}}^{2}(r)}, a_* = \frac{cJ_{\rm{r}}}{GM_{\rm{r}}^2},
\end{equation}
where $M_{\rm{r}}$ and $J_{\rm{r}}$ are the mass and angular momentum enclosed in the radius $r$, and $R_{\rm isco}$ is the stable circular orbit of the black hole, which is given by

\begin{equation}
    \begin{aligned}
    &R_{\rm isco}(M,a_*) = \frac{GM}{c^2}\{ 3+z_2-[(3-z_1)(3+z_1+2z_2)]^{1/2}\},\\
    &z_1 = 1+(1-a^{2}_{*})^{1/3}[(1+a_*)^{1/3}+(1-a_*)^{1/3}],\\
    &z_2 = (3a^{2}_{*}+z^2_1)^{1/2}.\\
    \end{aligned}
\end{equation}
Particles that circularise outside the ISCO will have sufficient angular momentum to
go into orbit around the black hole, and eventually accrete onto the black hole with a timescale of $t_{\rm acc}\sim2/\alpha \Omega_{\rm k}$, where $\alpha$ is the standard dimensionless viscosity parameter. 

The rate at which material falls back on the accretion disc depends on both the density profile of the star and its angular velocity profile. Here we adopt a widely used stellar model (model 16T1 of Woosley \& Heger 2006 \citep{Woosley2006ApJ...637..914W}), where the progenitor star has mass $14M_{\odot}$ and radius $5.18\times10^{10}$ cm. According to the most recent simulations, a rapidly rotating massive star would become highly stratified at the end of the evolution \citep{Hirschi2004}. For the purpose of this work, we use a step function

\begin{equation}
    \Omega(r) = \left\{
    \begin{aligned}
       0.0035~\rm rad~s^{-1}, \quad & 0 < r \leq 8.8 \times 10^{9} \rm{cm},\\
       0.0023~\rm rad~s^{-1}, \quad & 8.8 \times 10^{9} \rm{cm} < r \leq 5.18 \times 10^{10} \rm{cm},
    \end{aligned}
    \right.
\end{equation}
to describe the angular velocity profile, such that material in the star is divided into two parts: stellar core and envelope. Without the influence of the companion, the inner core ($r< 3.6 \times 10^{9} \rm{cm}$) would directly form a black hole with a mass of $\sim5 M_{\odot}$. Here we assume the compact companion has a mass of $\sim3M_{\odot}$, and its merger with the inner core will produce a BH with a mass of $\sim 8 M_{\odot}$. 
We note that the detailed merging process and the precise properties of the nascent black hole need to be calculated by numerical simulation. Our treatment here is a very rough estimation, but will not influence the main conclusion. 

The outer core would form an accretion disk within $t_{\rm eq}\sim100\rm s [(1+z)/3]$ s, and accrete onto the black hole with $t_{\rm acc} \ll t_{\rm eq}$, which explains the prompt duration of GRB 220711B. After the prompt stage, the mass and spin of the BH increases to $\sim13 M_{\odot}$ and $\sim0.9$. Part of the envelop ($8.8 \times 10^{9} \rm{cm} < r \leq 2.1 \times 10^{10} \rm{cm}$) would fall back and form an accretion disk within $t_{\rm eq}\sim100 [(1+z)/3]-500 [(1+z)/3]$ s, and accrete onto the black hole with $t_{\rm acc} \ll t_{\rm eq}$. The spin axis of the nascent BH is mainly determined by the orbital angular momentum, which could be misaligned with the spin axes of the outer core and envelope of the the progenitor star. Such a spin inconsistency causes the misalignment in the spin axis of the BH and the angular momentum axis of the BH-disk system, so that the dragging of the inertial frame (frame dragging) produced by a Kerr black hole drives the accretion disk to precess. This effect is known as Lense-Thirring (LT) precession \citep{1918PhyZ...19..156L}. The precession period $P$ can be estimated as \citep{Wilkins1972}

\begin{equation}
P = 2\pi(1+z)/\Omega_{\rm{LT}},
\end{equation}
where $\Omega_{\rm{LT}}$ is the precession angular velocity given by

\begin{equation}
\Omega_{\rm{LT}}(R)= \frac{2G}{c^2} \frac{J_{*}}{R^3},
\end{equation}
where $J_{*}=a_{*} G M_{\rm BH}^2/c$ is the BH angular momentum. 
The combination of the LT torque and the internal viscosity leads to a warped disk due to the Bardeen–Petterson effect \citep{Bardeen1975} (hereafter BP). The inner disk tends to align with the BH spin, while the outer region tends to remain in the original orientation. The transition radius between the two regimes is known as the Bardeen–Petterson radius $R_{\rm BP}$, which is the radius at which the warping propagation timescale $t_{\nu_2} = R^2/\nu_2$ equals the local forced precession rate $P/(1+z)$, where  $\nu_2$ is the viscosity associated with the vertical shear motion describing the diffusion of warping distortion through the disk \citep{Scheuer1996}. For such a warped disk, the centrifugally driven wind by the magnetic field threading the disk would collimate the GRB jet, making it precess with the angle and period defined from the launching site. This launching site is most likely at $R_{\rm BP}$, where the disk inclination angle changes significantly. A similar proposal was introduced to study other BH accretion systems \citep{Begelman2006}. For a BH with mass $\sim13 M_{\odot}$ and spin $\sim0.9$, we have $P \simeq 50[(1+z)/3]$ s at $R_{\rm BP}=42R_{\rm{g}}$ ($R_{\rm g}=G M_{\rm BH}/c^2$ is the gravity radius), which well explains the QPO found in the X-ray flares of GRB 220711B. The BP radius $R_{\rm BP}$, and therefore the precession period $P$, changes with the disk accretion rate. Assuming the BP radius is also the location where the disk angular momentum $J_d(R_{\rm BP}) \simeq M_{\rm d}(R_{\rm BP}) \sqrt{G M_{\rm BH} R_{\rm BP}} $ is comparable with the BH angular momentum $J_*$. The disk accretion rate depends on the mass and accretion timescale at $R_{\rm BP}$ as $\dot{M}=M_{\rm d}(R_{\rm BP}) /t_{\rm acc}(R_{\rm BP})$, where $t_{\rm acc} (R_{\rm BP}) = R_{\rm BP}^2/\nu_1$ and $\nu_1$ is the standard shear viscosity in a flat disk. The relation between $\nu_1$ and $\nu_2$ was obtained analytically by \cite{Ogilvie1999} as $\nu_2/\nu_1=2(1+7 \alpha^2)/(4\alpha^2+\alpha^4)$, where $\alpha$ is the viscosity parameter. One would thus find that the precession period $P\propto \dot{M}^{-6/7}$. As a result of a decrease of density for the stellar model, the decay of the fall-back rate (consequently a decay of the accretion rate) will cause a continuous increase of the precession period. Assuming that $\dot{M}\propto t^{-\lambda}$, where $\lambda$ is the 
decay index of the accretion rate, it is expected that $P\propto t^{6\lambda/7}$. From the observational point of view, the increase of the precession period could be manifested through a gradual widening of the X-ray flares. For the data of GRB 220711B, the duration of the X-ray flares indeed gradually increases with time as $\Delta T\propto t^{0.72}$, which is well consistent with the theoretical expectation with $\lambda\sim0.84$.

During the prompt stage, the accretion rate will be 2-3 orders of magnitude higher than that in the X-ray flares phase. Therefore, the LT precession period in the prompt phase will be $P\sim 0.5[(1+z)/3]$ s, which is hard to resolve due to the mixture with other instability timescales from the disk and jet. A precessing jet will periodically inject kinetic energy into the external shocks, which is equivalent to periodically launched sub-jets \citep{Huang2021}. In this scenario, once the equivalent sub-jets have a mass distribution with the bulk-motion Lorentz factor approximated as $M(>\gamma) \propto \gamma^{-2.45}$, the shallow decay behavior in the X-ray light curve of GRB 220711B could be well interpreted.

\section{Conclusion and Discussion}
\label{discussion}

Despite decades of research, our understanding of the progenitors and central engines of GRBs is still limited due to the lack of direct observational constraints. It has long been proposed that the observed temporal characteristics of GRB prompt and afterglow emission mirror the temporal behavior of the central engine, potentially offering insights into the progenitor. In this work, we conduct a comprehensive multi-wavelength data analysis on a peculiar gamma-ray burst, GRB 220711B. 

Based on its observational properties, GRB 220711B can be divided into three distinct emission episodes: a prompt emission phase with multiple peaks and no evident QPO; an early X-ray afterglow phase featuring a clear QPO signal; and a late-time X-ray afterglow with a typical decay index and jet break time. The prompt emission and X-ray afterglow of GRB 220711B largely fit within the traditional GRB framework. Incorporating the jet precession into our model, we propose a physical interpretation for this GRB: it likely originated from a binary system consisting of a typical GRB progenitor and a compact star. The compact star merged with the inner core of the progenitor, forming a BH and triggering the collapse. The prompt emission arose from accretion of the outer core, while the early afterglow, which includes the QPO, resulted from accretion of the progenitor’s envelope into a BH with a misaligned spin axis. 

So far in the literature, no GRB has been found to have a significant QPO signal. We systematically searched all GRB afterglow observed by XRT for QPO signals. The XRT light curves (between 2005 February and 2022 August) were downloaded from the public data available in the Swift archive (\url{https://www.swift.ac.uk/xrt_curves/}). We focus on the light curves consisting of multiple flares. By visual inspection, we first screen sources with more than three consecutive X-ray flares, whose duration are of the same order of magnitude. Besides GRB 220711B, 20 GRBs are selected: GRB 050730, GRB 060111A, GRB 060210, GRB 070129, GRB 071031, GRB 100212A, GRB 100728A, GRB 111123A, GRB 130514A, GRB 130606A, GRB 140304A, GRB 140506A, GRB 140817A, GRB 150323C, GRB 170714A, GRB 170810A, GRB 171212A, GRB 180620A, GRB 190926A, and GRB 210820A. We apply both WWZ and LSP methods to the data of these GRBs, but no significant QPO signal was found anywhere. The scarcity of QPO samples may be attributed to the rarity of progenitors for 220711B-like GRBs. On the other hand, in the case of jet precession, certain conditions are required for a significant QPO signal to appear in the light curve: first, the precession angle should be large relative to the jet opening angle, which requires that the misalignment in the spin axis of the BH and the angular momentum axis of the BH-disk system is large enough; secondly, the precession period is required to be smaller than the total emission duration, but the time scale for BH-disk restoring alignment should be equal to or longer than the total emission duration (the alignment effect tends to weaken the significance of such a QPO); finally, the signals containing QPO modulation should not be submerged by other signals, such as the tail emission of the prompt phase or the early emission from the external shock. This is why GRB 220711B is a rare case, and thus provides a good chance for investigating the properties of a GRB progenitor.

In addition to the precession model, alternative models may also account for the distinctive characteristics of GRB 220711B with fine-tuning conditions. For instance, \cite{Yamazaki2004} suggested that the jet of a GRB consists of multiple subjets, and the off-axis subjets could produce multiple consecutive flares. In this case, the duration of each flare is generally stochastic, and the emergence of QPO is feasible in theory but remains an unlikely coincidence.


Finally, we would like to point out that, although various methods have yielded relatively significant QPO features, the power spectrum analysis method does not show strong peaks around $0.02$ Hz. This can be ascribed to two potential reasons: 1) the power spectra analysis operates under the presumption that the bins in a periodogram conform to an $\chi^2$ distribution, thereby allowing a Whittle likelihood to be applied. However, this assumption is only accurate in the context of infinitely long time series that are characterized by homoscedastic stationary Gaussian data \citep{hubner2022searching}. Therefore, in the situations where the total sampling duration is limited, such as the cases when fast transients like GRBs are considered, the power spectral analysis method may demonstrate reduced effectiveness in detecting low-frequency signals \citep{Huppenkothen2013,Hübner2022a}; 2) We find that the duration of each X-ray flare gradually increases with time, which further reduces the significance of the QPO frequency in the power spectrum. 

Nevertheless, the unsuccessful power spectrum analysis to some extent suggests the possibility, albeit very low, that the similarity in timescales of the early X-ray flares in GRB 220711B could be a mere coincidence, resulting in a false appearance of a QPO signal. If sources similar to GRB 220711B are detected again in the future, particularly those with precession periods significantly shorter than the observation duration, it is hoped that QPO signals will also become prominent in power spectrum analyses. These sources will further support our hypothesis and offer new insights into studying the central engines and progenitors of gamma-ray bursts.

\begin{acknowledgments}

This work made use of the data from {\it Fermi} and {\it Swift}. This work is supported by the National Natural Science Foundation of China (Projects: 12021003,U2038107), the science research grants from the China Manned Space Project with NO. CMS-CSST-2021-A13, CMS-CSST-2021-B11, and the Strategic Priority Research Program on Space Science (Grant No.XDA15360000) of the Chinese Academy of Sciences. DAK acknowledges support from Spanish National Research Project RTI2018-098104-J-I00 (GRBPhot). During the revision process of this paper, D.A. Kann unfortunately passed away. Special thanks are extended to him for his contributions to this work, and this paper is dedicated as a tribute to his memory.

\end{acknowledgments}

%

\vspace{5mm}
\facilities{HST(STIS), Swift(XRT and UVOT), AAVSO, CTIO:1.3m,
CTIO:1.5m,CXO}


\software{astropy \citep{2013A&A...558A..33A,2018AJ....156..123A},  
          Cloudy \citep{2013RMxAA..49..137F}, 
          Source Extractor \citep{1996A&AS..117..393B}
          }




\bibliography{maintext}{}
\bibliographystyle{aasjournal}

\end{document}